\documentclass[style/psfig]{article}

\usepackage{graphicx}
\usepackage{setspace}
\usepackage[margin=10pt,font=small,labelfont=bf, labelsep=endash]{caption}
\usepackage[ruled,boxed]{algorithm2e}
\usepackage{amsmath}			%%% EWM: Needed for \align{} environment

\def \p    {\prime}
\def \pp   {{\prime\prime}}
\def \ppp  {{\prime\prime\prime}}
\def \pppp {{\prime\prime\prime\prime}}

\def \eps {\epsilon}

\def \ra {\rightarrow}
\def \oo {\infty}
\def \half {\frac{1}{2}}

\def \fns {\footnotesize}

\begin{document}

%
%
%--  Alphabetical Equation labelling [eg. (1a) (1b) (1c) ... ]
%
%--    Commands:   \abceqnbeg, \abceqnend
%--    Use in the following manner:
%--      \abceqnbeg
%--      \begin{equation} ... \end{equation}
%--      \begin{equation} ... \end{equation} ...
%--      \abceqnend
%
 \newcounter{abceqn}
 \def \abceqnbeg {\setcounter{abceqn}{\value{equation}}\addtocounter{abceqn}{1}\setcounter{equation}{0}\def \theequation{\arabic{abceqn}\alph{equation}}}
 \def \abceqnend {\def \theequation{\arabic{equation}}\setcounter{equation}{\value{abceqn}}}
%
%
%--  Alphabetical figure labelling [fig. (1a) (1b) (1c) ... ]
%
%--    Commands:   \abcfigbeg, \abcfigend
%--    Use in the following manner:
%--      \abcfigbeg
%--      \begin{figure} ... \end{figure}
%--      \begin{figure} ... \end{figure} ...
%--      \abcfigend
%
 \newcounter{abcfig}
 \def \abcfigbeg {\setcounter{abcfig}{\value{figure}}
                  \addtocounter{abcfig}{1}  \setcounter{figure}{0}
                  \def \thefigure{\arabic{abcfig}\alph{figure}} }
 \def \abcfigend {\def \thefigure{\arabic{figure}}
                  \setcounter{figure}{\value{abcfig}}}
 \newcounter{appabc}
 \def \appabcbeg {\setcounter{appabc}{\value{equation}}
                  \addtocounter{appabc}{1}  \setcounter{equation}{0}
                  \def \theequation{\arabic{appabc}\alph{equation}} }
 \def \appabcend {\def \theequation{\arabic{equation}}
                  \setcounter{equation}{\value{appabc}}}

\centerline{\bf Long's Vortex Revisited}
\bigskip

\bigskip

\centerline{4 Mar 2013}
\bigskip
\centerline{Sih-Tsan~Lee and Ernst~W.~Mayer${}^*$}
%%%\centerline{\fns 10190 Parkwood Dr.~\#1, Cupertino, CA 95014}
\centerline{\fns ${}^*$ To whom correspondence should be addressed: {\em ewmayer@aol.com}.}
\bigskip

%\section*{Abstract}
\begin{abstract}

The conical self-similar vortex solution of Long (1961) is reconsidered, with a view toward understanding what, if any, relationship exists between Long's solution and the more-recent similarity solutions of Mayer and Powell (1992), which are a rotational-flow analogue of the Falkner-Skan boundary-layer flows, describing a self-similar axisymmetric vortex embedded in an external stream whose axial velocity varies as a power law in the axial ($z$) coordinate, with $\phi=r/z^n$ being the radial similarity coordinate and $n$ the core growth rate parameter. We show that, when certain ostensible differences in the formulations and radial scalings are properly accounted for, the Long and Mayer-Powell flows in fact satisfy the same system of coupled ordinary differential equations, subject to different kinds of outer-boundary conditions, and with Long's equations a special case corresponding to conical vortex core growth, $n=1$ with outer axial velocity field decelerating in a $z^{-1}$ fashion, which implies a severe adverse pressure gradient. For pressure gradients this adverse Mayer and Powell were unable to find any leading-edge-type vortex flow solutions which satisfy a basic physicality criterion based on monotonicity of the total-pressure profile of the flow, and it is shown that Long's solutions also violate this criterion, in an extreme fashion. Despite their apparent nonphysicality, the fact that Long's solutions fit into a more general similarity framework means that nonconical analogues of these flows should exist. The far-field asymptotics of these generalized solutions are derived and used as the basis for a hybrid spectral-numerical solution of the generalized similarity equations, which reveal the existence of solutions for more modestly adverse pressure gradients than those in Long's case, and which do satisfy the above physicality criterion.

\end{abstract}

%%%%%%%%%%%%%%%%%%%%%%%%%%%%%%
\section{Outline}
\label{sect:outline}

A brief outline of the paper is as follows: We first review the Hall and Mayer-Powell leading-edge-type and Long "tornado-like" similarity solutions. We then show how both of the different similarity procedures used for these disparate flow types can be reduced to a single unified framework accommodating a continuum of growth-rate-parameterized solutions, in which what we dub "Long-type" solutions satisfy the same similarity equations derived by Mayer and Powell but with a far-radial-field at infinity. The particular flows described by Long are a special conical-vortex-growth case of the general solutions.  We next derive asymptotic series expansions describing the behavior of the generalized Long-type flows in the far radial field, and describe a novel hybrid spectral-numerical procedure for solving the corresponding full similarity equations in the inner viscous core region which automatically includes the matching to the appropriate far-field asymptotic solution, by way of treating the unknown far-field series coefficient as an eigenvalue of the problem. The numerical results show that for less-adverse pressure gradients than those in Long's case (and thus having viscous core growth rates less than conical), these full-field solutions do satisfy the above physicality criterion based on total-pressure. The asymptotic analysis of the far-field behavior of the generalized solutions also shows that there is a close correlation between the onset of nonphysicality according to the total-pressure criterion and singular behavior in the coefficients of the asymptotic series. The singularity is only apparent when one considers a continuum of solutions with different values of the growth-rate parameter, and thus would not be apparent if a single similarity law (e.g.~conical growth, as in Long's flows) is considered.

\section{Introduction}
\label{sect:intro}

In this study, we consider an idealized model of axisymmetric vortices with slender cores in an incompressible fluid having infinite extent. Under these assumptions, two kinds of vortex flows are frequently studied. The first kind is a leading-edge-type vortex with nonzero axial velocity excess over that in a ``far field'' taken to be sufficiently outside the immediate vicinity of the viscous core but not extending to infinity; herein we will refer to such flows as ``Hall-type vortices'' for simplicity. The second kind consists of a viscous swirling core embedded in a potential vortex, i.e. having circulation $\Gamma = vr$ tending to a constant in the far field, and all velocity components decaying to zero far from the axis; in the context of the present study we shall refer to these type of solution of the governing equations as ``Long-type vortices.''

\cite{Hall} proposed a similarity-solution model for vortices formed by the rolling up of the shear layer produced at the leading edge of a lifting delta wing, which divided the vortex core into two regions: an inviscid rotational conically self-similar outer core, with viscous effects appreciable only in a nonconical inner subcore. This approach yields an exact solution for the conical outer region with velocity components ($u$ = radial, $v$ = azimuthal, $w$ = axial) of the form
\begin{equation}
u=-\sigma\half W_e\ \phi,\qquad v=\left[V_e^2 - \sigma^2 W_e^2 \log\left(\frac{\phi}{\phi_e}\right)\right]^\half,\qquad w=W_e\left[1 - \sigma \log\left(\frac{\phi}{\phi_e}\right)\right],
\label{hall_outer}
\end{equation}
where $\phi= r/z$, $\sigma = \sqrt{1+2V_e^2/W_e^2} - 1$ and $\phi_e$ is an outer-boundary-condition location whose value can be chosen - the only obvious constraint is that it should be sufficiently far from the rotational axis that viscous effects are negligible there,~- a requirement which can be checked a posteriori~- and $V_e$ and $W_e$ are the azimuthal and axial velocity values (also user-specified) at this outer edge location. For the inner viscous core, Hall was able to find a solution by making scaling assumptions of the usual boundary-layer type, subject to two key additional assumptions: (i) the axial velocity within the viscous subcore is nearly constant (which is not justified physically, based on the large axial-velocity variations seen in such flows in experimental studies such as the contemporaneous one of \cite{Earnshaw}, and (ii) the radial velocity has a far-field near-linear velocity profile which matches the inviscid outer core solution. With these additional assumptions, Hall was able to reduce the governing equations to an analytically tractable form. The inner-core solutions obtained this way are in fairly good agreement with Earnshaw's experimental data in the outer part of the viscous subcore region, but give predictions for the inner core (the region of greatest interest) which only agree with the experimental data in a rough qualitative sense. Hall's approach additionally made it difficult to examine the solution behavior in the limit of vanishing viscosity, and utilized an unsatisfactory ad hoc matching procedure between the inner and outer solutions. Soon thereafter, \cite{StewartsonHall} improved the inner subcore solution via a proper asymptotic matching approach; the actual results, however, were little different from Hall's original work. Fundamentally, due to the nature of the above outer core flow, the solutions can only be extended a finite distance from the vortex axis irrespective of which matching technique one uses. This is perhaps not terribly troubling from the standpoint of understanding leading-edge-type vortex cores since such flows naturally have a limited radial extent (since the vorticity is introduced at a finite radial distance by the shear layer rolling up from the leading edge), but it does beg the question of whether there is any kind of flow model of this type which leads to solutions which can be extended radially to infinity, as those for wake-like and potential vortices can.

\cite{MayerPowell1} generalized Hall's approach by allowing for variation of axial velocity in the far field of power-law form $W(z)\sim z^m$. They made the added simplification of using a single similarity variable for the entire core region in lieu of the separate ``inner'' and ``outer'' core approach used by Hall. A straightforward similarity analysis showed that the viscous core region has an axial growth rate described by a parameter $n$ which appears by way of the radial similarity variable $\phi = r/(\eps z^n)$ and is coupled to the power-law parameter that describes the axial variation of the far field via the simple relation $n = \frac{1-m}{2}$. (The power-law scalings for these flows are in fact identical to those of the well-known Falkner-Skan boundary-layer flows, which analogously generalize the famous solution of Blasius (1908) for the viscous boundary layer formed by flow over a flat-plate with constant far-field conditions.) They thus obtained a set of similarity equations describing a generalized family of Hall-type flows.

The equations for $m=0$, i.e.~with no axial velocity or pressure gradient and with nonconical self-similarity, admit an analytical solution in the outer core region which is identical to Hall's outer solution~(\ref{hall_outer}), but with the radial velocity $u$ multiplied by the growth rate parameter $n$. Owing to the single similarity variable, the full viscous solution can be obtained numerically to any desired accuracy without requiring a tedious matching process between the inner and outer core regions, and the $m=0$ solutions agree quite well with the high-quality experimental data of \cite{VerhaagenRansbeeck} for incompressible flow over a delta wing. For $m$ decreasing from 0, which corresponds to an increasingly adverse axial pressure gradient of the outer flow, the axial velocity profile of the solutions shows an increasingly pronounced retardation in the viscous region near the vortex axis, flow behavior which is reminiscent (within the obvious limitations of the similarity approach) of the large-scale flow disruptions seen in the phenomenon of vortex breakdown in real-world flows (cf.~\cite{Sarpkaya}).

Note that in both the original Hall-type formulation and its generalizations, since one is specifying boundary conditions at a finite distance from the vortex axis, unless one manages to eliminate the pressure from the formulation, this includes specifying a pressure value at the outer boundary. For $m=0$ the pressure appears in the similarity equations only via its derivative, hence this outer-boundary value is irrelevant in the sense that neither the velocity profiles nor the degree of pressure change between the outer flow and the vortex core depend on the chosen boundary value. However, for nonzero values of $m$ an undifferentiated pressure term appears in the similarity equations, as well, which means that the chosen value of edge pressure has a deterministic effect on the solution properties. Mayer and Powell dealt with this issue by showing that there is a particular choice of edge pressure such that the resulting distributions of total pressure $p+\half(v^2+w^2)$ (in terms of the similarity scalings, the radial velocity $u$ is negligible and drops out of this expression) are asymptotically constant and decrease to a minimum on the axis, which is physically realistic based on behavior of real flows. These finite-boundary-distance-related issues are absent in flow models in radially unbounded domains such as the Long-type flows we shall describe next, even though (as we will also show) the governing similarity equations are the same.

Dating back to virtually the same time period as the leading-edge-vortex work of Hall, the family of conically self-similar rotational solutions of the steady incompressible Navier-Stokes equations discovered by \cite{Long} and collectively referred to as ``Long's vortex'' is commonly used as an idealized model of tornado-like flows~(\cite{BurggrafFoster,ShternHussain}) and as a reference flow for studies of swirling-flow stability~(\cite{FosterDuck,FosterSmith,FosterJacqmin,KhorramiTrivedi,ArdalanDraperFoster}). Because of this wide interest, one would assume that the basic flows themselves have been subjected to close scrutiny, found to be physically reasonable according to some rational set of criteria, and the details of their scalings understood and placed in some clear relation to other boundary-layer-like flows, of both the swirling and non-swirling variety. We find that this is not the case. Some important points left unaddressed in this regard in the extant literature are:
\begin{enumerate}
\item In his derivation, Long used a ``boundary-layer-like'' (his words) scaling for the radial similarity variable of $y\sim r/(\eps_L z)$, where $\eps_L = \mbox{Re}^{-1}$ and the Reynolds number is based on outer-flow circulation. Based on this definition, the radius of the core grows linearly with axial distance downstream of a singular point at the apex of the vortex, and is inversely proportional to Reynolds number. Compared to the ${\cal O}(\mbox{Re}^{-\half})$ thickness growth rate of other well-known boundary-layer flows, the scaling used by Long is unusual. (In fact it is precisely this odd scaling that first led us to reinvestigate Long's formulation.)
\item Perhaps the most prominent feature of Long's solutions is their nonuniqueness with respect to a characteristic parameter $J$, the so-called ``flow force,'' equivalent to the total flux of axial momentum through any plane perpendicular to the vortex axis. For any given value of $J$ in a particular range, there exist two sets of solutions of the similarity equations, which Foster and Duck (and subsequent authors) labeled type I and type II flows, and which are mainly differentiated by their axial velocity components: the type I flows exhibit regions of reversed axial flow, whereas the type II flows do not. We shall show that the solutions can in fact be alternately and uniquely parameterized by their axis values of axial velocity, and thus that the apparent nonuniqueness is strictly a result of Long's particular choice of parameterization. However, from a physical standpoint, no plausible physical mechanism has been advanced for the reversed flows.
\item At large values of the flow force, \cite{FosterSmith} found flow variations on the reversed-flow (type II) solution branch to become intensely concentrated in an asymptotically narrow annulus about $r = 4\pi^2 J/(\rho\Gamma^2)$ where axial and azimuthal velocities are large, a flow pattern they dub a ``ring-jet.'' Peculiar as such solutions are, no questions were raised about their physical plausibility.
\item In Long's solutions, the static pressure drops monotonically from a constant value in the far field to a minimum on the axis, as expected on physical grounds for a swirling flow. However, to the best of our knowledge, no total-pressure plot for any of these flows has previously appeared in the published literature. If one neglects the temperature change caused by viscous dissipation (as is justified in the incompressible-flow limit), the total pressure represents the variation of total flow enthalpy. As there is no energy source embedded in the flow, one expects the total pressure to reach a minimum on the vortex axis (where the cumulative dissipation is greatest) and to increase toward a constant value in the far field. This is a minimum requirement for such a flow to be physically plausible. We find that none of Long's solutions satisfy this simple total-pressure-based physicality criterion, and that it is violated particularly badly by the type I solutions, those with a strictly jet-like axial flow profile
\footnote{
A reviewer has objected to this criterion as being unjustified under the assumptions used in the flow model, and indeed, based on strictly icnompressible flow, it is, as there is no energy equation, no second law of thermodynamics, and hence no dissipative ``losses" which should be reflected in added constraints on the solutions. But in fact theoretical fluid mechanics abounds with such apparent paradoxes and solution selections ``unjustified by the model". A classical example is the idealized 2-D model of the circular spinning cylinder used by such luminaries as Rayleigh to illustrate crucial concepts related to circulation and lift as embodied by the Magnus effect. If the cylinder is truly of a perfectly circular cross-section with a perfectly smooth surface and the fluid is truly inviscid there is of course no way for the fluid to ``know" whether the cylinder is spinning or not. The ``unjustified" thing done is to invoke the viscous vorticity transfer known to occur in real fluids as a result of the no-slip condition. Does this mean we should ignore Rayleigh's work and conclusions?

We feel it important, even in the context of a highly idealized athematical model of the physics of interest, to emphasize those solutions of the model equations which appear to be of greatest relevance to phenomena observed in the real world. In the present case, we use knowledge of total-pressure profiles seen for similar flows in actual experiments to restrict our focus. The generalized similarity formulation certainly admits of a similarly wide variety of ``heretofore unseen in the laboratory or in nature" solutions as the does the original conical Long's formulation. Readers interested in studying these wilder regions of the solution space are welcome to do so to their hearts' content.
}.
\end{enumerate}

Due to these unresolved issues with respect to the Long's flows, we decided to undertake a thorough review of the basic formulation and solutions for the Long-type vortex flows, not only to to be able to say something about their physical plausibility (or lack thereof), but also in an effort to try to place them in some kind of context among other well-known vortical-flow similarity models. The gist of our findings, and an outline of the paper, is as follows:

In \S\ref{sect:formulation} we show that the similarity equations derived by Long are in fact the same as those obtained via more-orthodox boundary layer scalings. A corollary of this analysis is that, when properly scaled, the similarity laws and equations governing Long's vortex are precisely the same as those derived by \cite{MayerPowell1} in their investigation of leading-edge-type swirling flows, but subject to a different set of far-field boundary conditions. Placing Long's flows in this context also makes clear why reversed-flow solutions might be expected to exist: conical self-similarity, as assumed {\it a priori} by Long, is only possible if the outer axial flow decelerates in a $z^{-1}$ fashion, which implies a highly adverse axial pressure gradient, in fact one which is much more severe than the most-adverse case for which the aforementioned authors were able to find physically reasonable leading-edge-type solutions, and similar nonphysicality is evinced by Long's solutions as well, once one examines their total pressure profiles. Fortunately, all is not lost. Since the Long's flows are solutions of a special case (having $m=-1$) of the Mayer-Powell similarity equations, there exists the possibility that there are generalized Long's flows of nonconical form, most especially ones at less-adverse axial pressure gradients ($m>-1$) and hopefully having more-reasonable total pressure distributions. The derivation of these generalized (nonconical) analogues of Long's similarity equations concludes the section.

Prior to attempting a numerical solution of the generalized similarity equations, it is worth examining (and in fact necessary in the context of the spectral-numerical approach used here) the asymptotic behavior of solutions in the far field, and this is the subject of \S\ref{sect:asymp}. It is shown that solutions of the equations for which all velocities decay at radial infinity are only possible if the pressure gradient is adverse ($m<0$) in terms of the Falkner-Skan-type similarity parameter governing the flow.  General asymptotic series suitable for describing the dependent variables (stream function and circulation) are developed, which reduce to the well-known far-field series governing Long's solutions in the special case of conical self-similarity, but which also prove suitable for nonconical flows.

%%%For the sake of completeness, the fundamental integrals expressing axial and angular momentum flux in the axial direction are examined for the generalized solutions. It is found that one branch of solutions that satisfy the total-pressure criterion also has constant axial momentum flux, and that none has constant angular momentum flux, but that nonconstancy of these momentum-flux integrals is not inimical to physical realizability.

Information about the viscous core region appears in the asymptotic far-field series as an undetermined coefficient in terms of which the coefficients of all the higher-order terms can be expressed, and which in a classical matched-asymptotic-expansion approach would be successively approximated by matching to a functional hierarchy which approximates the viscous inner solution. We adopt a rather different approach to the matching by treating the undetermined coefficient as an eigenvalue to be found in a numerical solution spanning the entire flow field, which is constructed (using the preceding far-field series expansion) to have the proper asymptotic behavior in the far radial field, assuming the ``matching constant'' yielded by the numerical solution procedure converges as the numerical resolution is increased. This procedure is described in \S\ref{sect:numerics}, as is the high-accuracy iterative spectral method used to solve the resulting system of coupled nonlinear ordinary differential equations containing the necessary eigenvalue term.

The character of the generalized flows is the subject of \S\ref{sect:solutions}. We begin by reviewing the properties of the classical conical case considered by Long in \S\ref{sect:neq1}. As previously mentioned, many aspects of these solutions are well-studied, and we use our numerical results and the accompanying plots mainly to illustrate key qualitative aspects of the solutions. The discussion concludes by presenting the total-pressure profiles for a flow which typifies the Type I solution branch to demonstrate the likely nonphysicality of the conically-similar solutions.

The characteristics of the generalized Long-type flows are the subject of \S\ref{sect:gensolutions}. An interesting property of the generalized similarity equations is that the form of the higher-order coefficients of the corresponding asymptotic series derived in \S\ref{sect:asymp} permits circumstances under which these coefficients blow up due to a vanishing denominator. This can only happen if the flow in question has a growth rate which is less than conical, and some examples where this singular behavior actually does occur are detailed. Perhaps tellingly, the locations in the parameter space where such behavior occurs correlate closely with the onset of near-axis overshoots of the total pressure, so this ``coefficients crisis'' may serve as a useful proxy for the onset of nonphysicality as based on this criterion. In \S\ref{sect:long2hall} we conclude by explicitly demonstrating how one may start with a generalized Long-type solution, cut it off at finite radius, and smoothly transition to a Hall-type solution at the same flow parameters by varying the outer-edge pressure value, assuming one is in a region of the parameter space (defined by growth-rate parameter $n$ and core axial velocity) for which both types of solutions exist.

%%%%%%%%%%%%%%%%%%%%%%%%%%%%%%
\section{Governing Equations and Similarity Formulation}
\label{sect:formulation}

The governing equations are the usual time-invariant incompressible axisymmetric Navier-Stokes equations. The only salient difference in this regard between the Hall and Long-type formulations is that in the former one typically uses a characteristic axial velocity with which to nondimensionalize the flow variables and on which to base the Reynolds number, whereas in the latter one uses the asymptotic value of far-field circulation for this purpose. In his formulation Long began with the second (circulation-based) definition of the Reynolds number and used that to define a small parameter $\eps_L$ and radial similarity variable $y$, with
$y$ assumed to be of order unity in the viscous vortex core:
\begin{equation}
\eps_L := \mbox{Re}^{-1},\qquad y := \eps_L^{-1} r/(\sqrt{2} z),
\label{long_eps}
\end{equation}
along with a nondimensionalized stream function $f(y)$ and circulation $g(y)$, also taken to be of order unity. Long claimed the above definition of small parameter and scaling of the radial similarity variable to be ``boundary-layer-like,'' but in a standard boundary layer treatment one instead uses $\eps = \mbox{Re}^{-1/2}.$ However, it is easily verified that Long's eventual similarity equations in fact express the same balance of terms which results from a more conventional boundary layer approach. In fact, the potential confusion here arises because Long's scalings in effect correspond to the use of a characteristic {\em radial} (i.e.~transverse) velocity in the nondimensionalization procedure and definition of the Reynolds number. (Whether the resulting flow solutions actual exhibit a radial velocity distribution which allows for a suitable reference radial velocity value to be defined is thus rendered moot.) With the further expedient of omission of the superfluous $\sqrt{2}$ factor introduced by Long into the denominator of the radial similarity variable, the resulting equations (still in terms of nondimensionalized velocities and pressure) are identical to the conical-form special case $n=-m=1$ of those obtained by \cite{MayerPowell1} based on the radial similarity coordinate $\phi = r/(\eps z^n)$:
\abceqnbeg
\begin{eqnarray}
u^\p+\frac{u}{\phi} &-& n\phi w^\p = 0,\qquad{\rm (continuity)}\label{massd}\\*
p^\p &=& \frac{v^2}{\phi},\qquad(r-{\rm momentum)}\label{rmomd}\\*
v^\pp-\left[u-n\phi w-\frac{1}{\phi}\right]v^\p &-& \left[mw+\frac{1}{\phi}\left(u+\frac{1}{\phi}\right)\right]v=0,\qquad(\theta-{\rm momentum)}\label{tmomd}\\*
w^\pp-\left[u-n\phi w-\frac{1}{\phi}\right]w^\p &+& nv^2-m\left(w^2+2p\right)=0,\qquad(z-{\rm momentum)}\label{zmomd}
\end{eqnarray}
\abceqnend
where we leave things in terms of both the velocity-field axial power-law parameter $m$ and the viscous boundary-layer growth-rate parameter $n$ for simplicity, even though these parameters are related via $m=1-2n$, the same relation as in the classical Falkner-Skan boundary layer flows.

For a leading-edge-type flow, where the spiral roll-up of the shear layer creates a radially finite vorticity field, suitable boundary conditions are
\begin{equation}
u=v=w^\p=0\ {\rm at}\ \phi=0;\qquad v=v_e, w=1, p=p_e=\half(1+v_e^2)\ {\rm at}\ \phi=\phi_e.
\label{bc_hall}
\end{equation}
The choice of ``outer edge'' location $\phi_e$ and edge swirl $v_e$ determine the strength of the resulting vortex. The edge pressure boundary condition (which is non-arbitrary for $m\ne0$ because in that case the pressure is nontrivially coupled to the velocities by way of its appearance in the axial momentum similarity equation) is obtained from assuming the flow to be irrotational in the far field ($\phi_e$ must be taken sufficiently large to ensure that this approximation is reasonable) and applying Bernoulli's law, with the total pressure taken as zero in the far field. The reason an analogous issue does not not arise for the Falkner-Skan boundary-layer flows is that in the latter, the pressure may be assumed constant in the transverse direction, whereas in a swirling flow this is not the case. Of course for $m\ne0$, solutions of the above equations may exist for other values of $p_e$ than that used above which differ nontrivially (i.e.~in more than just an arbitrary additive constant) from those using $p_e=\half(1+v_e^2)$, but as described by Mayer and Powell, none of them appear to satisfy the basic criterion of physicality we use here, namely that the total pressure $p_0=p+\half(v^2+w^2)$ tends monotonically toward a constant as $\phi\rightarrow p_e$.

For convenience, we will refer to solutions of the similarity equations subject to boundary conditions of the form~(\ref{bc_hall}) as {\em Hall-type} solutions. A second class of solutions, and the main topic of this paper, is represented by solutions which extend to infinity (when such exist), and for which all flow variables decay to zero in the radial far field, i.e.~subject to outer boundary conditions
\begin{equation}
u,v,w,p\rightarrow 0\ {\rm as}\ \phi\rightarrow\infty.
\label{bc_long}
\end{equation}
Solutions satisfying these latter boundary conditions will henceforth be referred to as {\em Long-type} solutions. Since all velocity components decay to zero for this latter class of flows, one cannot adjust the strength of the vortex in the same way on does for Hall-type flows, by fixing nonzero values of $v$ and $w$ at the outer boundary; rather, one must do something at the vortex axis, for instance setting the axial velocity there by replacing the inner boundary condition $w^\p(0)=0$ with $w(0)=w_{axis}$. (Alternatively, one might specify $v^\p$ on the vortex axis, in effect specifying the vortex strength by way of the axial component of vorticity there.)

%%%%%%%%%%%%%%%%%%%%%%%%%%%%%%
\subsection{On the parameterization of solutions}
\label{sect:param}

The family of flows first studied by Long have historically been parameterized via a quantity called the ``flow force,'' defined as
\begin{equation}
J := \frac{\rho\Gamma^2}{4\pi^2}M = \int_0^\infty(p-p_\infty+\half\rho w^2)r dr
\label{flow_force}
\end{equation}
and which is essentially the nondimensionalized axial momentum flux of the flow through a plane of constant $z$. The proportional parameter $M$ is the one most often used to parameterize solutions; for instance Foster and Duck found no solutions for $M<3.75$, and two distinct solutions for each value of $M$ greater than this critical value (cf.~figure~3 in their paper).  Foster and Duck refer to solutions on the upper and lower part of the flow-force curve as Type I and Type II flows, respectively; the Type I flows are distinguished by their larger axial velocity maxima, whereas the Type II flows (except for a very small portion of the parameter space near the minimum $M$ for which solutions exist) exhibit regions of reversed axial flow in their cores. The Type I flows also exhibit axial velocity deficits in their cores for small $M$, but the deficits become less pronounced with increasing flow-force, and for $M>4.71$ (as determined by the above authors), the axial velocity profiles become strictly monotone decreasing functions of the radial coordinate. The existence of two solutions for each value of $M$ greater than the critical one has been accepted as an indication of nonuniqueness of solutions, but this interpretation is a parameterization-dependent one: the same aforementioned figure used by Foster and Duck -- which shows a solution curve in a Cartesian coordinate system having $M$ as its abscissa and the core axial velocity $w(0)$ as ordinate -- demonstrates that solutions are in fact unique if instead parameterized by $w(0)$, and this is the parameterization we shall use here.

%%%%%%%%%%%%%%%%%%%%%%%%%%%%%%
\subsection{Stream-function/circulation formulation}
\label{sect:fg}

To show explicitly that equations~(\ref{massd}-d) contain Long's equations as a special case, and also to simplify the far-field asymptotic analysis of the generalized Long-type flows, we now reformulate the similarity equations in terms of a stream function $f$, chosen to automatically satisfy the continuity equation~(\ref{massd}), and a circulation function $g$, defined as follows:

\begin{equation}
u:=nf^\p-\frac{f}{\phi},\qquad v:=\frac{g}{\phi},\qquad w=\frac{f^\p}{\phi}.
\label{fg_def}
\end{equation}

For the case of conical flow ($n=1$) these definitions are equivalent to Long's, except that Long multiplies the right-hand sides of the above three formulae by factors of $1/(\sqrt{2}\eps)$ $1/(\sqrt{2}\eps^2)$ and $1/(2\eps^2)$, respectively, and Long's similarity variable is $1/(\sqrt{2}\eps)$ times our definition of $\phi$. Since these factors of $\sqrt{2}$ and $\eps$ are completely unnecessary we eschew them in our analysis, except for the purposes of comparison with previous results which make use of Long's definitions. In terms of our more-orthodox definition of the small parameter $\eps$, the key variables in our formulation (in terms of independent variable $\phi$) and Long's dependent variables $\tilde{f}$, $\tilde{g}$, $\tilde{s}$ and independent variable $y$ are related simply as

\begin{equation}
f(\phi)=f(y),\qquad g(\phi)=\frac{1}{\eps}g(y),\qquad p(\phi)=-\frac{1}{\eps}s(y),
\label{us_vs_them}
\end{equation}

where $s(y)$ is the scaled, dimensionless static-pressure-related variable in Long's notation.

Rewriting the three momentum equations in terms of the new variables, and also differentiating the $z$-momentum equation~(\ref{zmomd}) once and using the $r$-momentum equation~(\ref{rmomd}) to eliminate the pressure term from the result yields the following system:

\abceqnbeg
\begin{eqnarray}
\phi^3 p^\p&-&g^2=0,\label{sim_cooked1}\\*
\phi g^\pp+(f-1)g^\p&+&(n-1)f^\p g=0,\label{sim_cooked2}\\*
\phi^3 f^\pppp+(f-2)\phi^2 f^\ppp+\bigl[(4n-1)\phi f^\p&-&3(f-1)\bigr](\phi f^\pp-f^\p)+2\bigl[n\phi g^\p+(n-1)g\bigr]\phi g=0,\label{sim_cooked3}
\end{eqnarray}
\abceqnend

where, for the sake of subsequent analysis, we have also rewritten things entirely in terms of $n$, the viscous layer growth-rate parameter. Note that for $n=2$, the $f,g$ form of the $\theta$-momentum equation~(\ref{sim_cooked2}) can be integrated once:

\begin{equation}
\phi g^\p+(f-2)g=k_\theta,
\label{k_theta}
\end{equation}

where $k_\theta$ is the integration constant. However, this proves not terribly useful since in practice $n=2$ appears to be far beyond the range of $n$ in which the resulting flow solutions are physically reasonable. The pressure-eliminated $z$-momentum equation~(\ref{sim_cooked3}) can be exactly integrated for two particular values of $n$, namely $\half$ and 1, which can be seen by recasting (\ref{sim_cooked3}) in the following form:

\begin{equation}
\phi\Psi^\p+\frac{(n-2)}{n}\Psi\ =\ \frac{4(n-1)(2n-1)}{n}f^\p\left[\phi f^\p-\frac{1}{n}f+\frac{2(n-1)}{n^2}\right],
\label{psi_eqn}
\end{equation}

where $\Psi$ is defined via the rather unenlightening (and ultimately unimportant, except for this exercise in evaluation of integrability) relation

\begin{equation}
\Psi=\phi^2 f^\ppp-\frac{5n-2}{n}(\phi f^\pp+ff^\p)+\frac{(13n^2-14n+4)}{n^2}f^\p+\phi ff^\pp+(2n-1)\phi(f^\p)^2+n\phi g^2.
\label{psi_def}
\end{equation}

The only important aspect of the definition of $\Psi$ is that it makes it easy to see that for $n=\half$ and $n=1$ the right-hand side of equation~(\ref{psi_eqn}) vanishes identically, allowing the left-hand side to be integrated. For $n=\half$ (Blasius-like viscous layer growth, zero axial pressure gradient) integration of equation (\ref{sim_cooked3}) yields

\begin{equation}
\phi^2 f^\ppp+(f-1)(\phi f^\pp-f^\p)+\half\phi g^2=C\phi^3=0,
\label{int_half}
\end{equation}

where the integration constant $C$ can be shown to be zero by direct substitution of the definitions (\ref{fg_def}), along with the setting of $n=\half$, ($m=0$) into the primitive-variable $z$-momentum equation~(\ref{zmomd}).

For $n=1$ (conical self-similarity, Long's case) integration yields

\begin{equation}
\phi^2 f^\ppp+(f-3)\phi f^\pp+\bigl[\phi f^\p-3(f-1)\bigr]f^\p+\phi g^2=D\phi^3=0,
\label{int_one}
\end{equation}

where the integration constant $D$ can again be shown to be zero, although the analysis required to do so is bit less trivial than in the case $n=\half$ (cf.~\cite{LeeThesis}.)

The combination of equation~(\ref{sim_cooked2}) together with either~(\ref{sim_cooked3}), (\ref{int_half}) or (\ref{int_one}) constitutes a coupled system of two equations in two unknowns which can be solved as it stands, but to complete the comparison with Long's formulation we integrate (\ref{int_one}) once more, using that $g^2=\phi^3 p^\p$ to obtain

\begin{equation}
\phi f^\pp+(f-1)f^\p+\phi^3 p=E\phi^3=0,
\label{int_two}
\end{equation}

where the integration constant $E$ again must be zero due to the requirement that the nondimensional pressure $p$ vanish at infinity. We have thus recovered Long's equations.

In the general case, equations~(\ref{sim_cooked1}-c) constitute a seventh-order system, fourth-order in $f$, second-order in $g$ and first-order in $p$. For the case of the generalized Long-type solutions, i.e.~on a semi-infinite domain, we require all velocity components to vanish at infinity~-- as a consequence of this, the nondimensional pressure must vanish at infinity as well, except in the special case $n=\half$, where its far-field value is an arbitrary constant (hence can also be chosen zero with no loss of generality.) Thus, in terms of the dependent variables $f$, $g$ and $p$, suitable boundary conditions are

\begin{equation}
\phi=0: f=f^\p=g=0, f^\pp=w_{axis};\qquad \phi\ra\infty: f\sim f_{as}, g\sim g_{as}, p\ra 0,
\label{bc_gen}
\end{equation}

where $f_{as}$ and $g_{as}$ refer to the asymptotic form of these variables in the far field, which must be known (at least to leading order) for our high-accuracy numerical solutions method to work; the far-field asymptotics are the subject of the next section.

However, before getting into the details of the asymptotics and numerics, we first briefly discuss why Long's solutions fail the total-pressure criterion for physicality by way of comparison with the generalized Hall-type flows studied by Mayer and Powell, and thus further motivate an analogous generalization of the Long-type solutions. We first define the static and total pressure coefficients in a slightly generalized version of the typical fashion, which applies equally well to both finite and infinite-domain flows:

\begin{equation}
C_p:=\frac{p-p_{e}}{\half\rho W_e^2},\qquad C_{p_0}:=\frac{p_0-p_{0e}}{\half\rho W_e^2} .
\label{pcoeff}
\end{equation}

\begin{figure}
\begin{centering}
\includegraphics[scale=0.6]{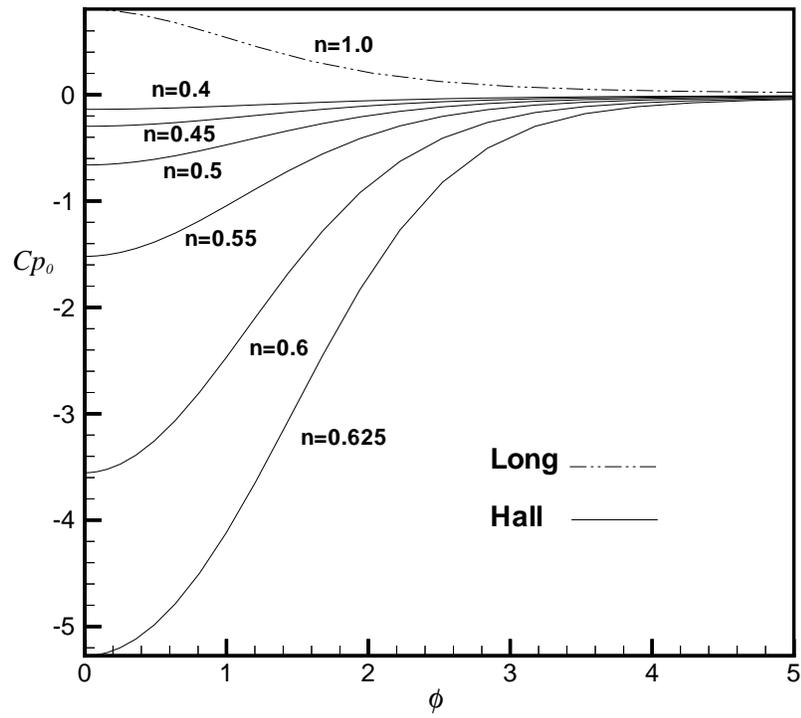}
\vspace{-0.2in}
\caption{Example total pressure coefficient distributions for Long's ($n=1.0$) and the generalized Hall-type solutions.}
\label{cp0_Long_Hall}
\end{centering}
\end{figure}

Here the subscript $e$ denotes some chosen ``outer edge'' location, which can be finite but need mainly be sufficiently large that the vorticity of the flow solution there be negligible compared to its maximum value in the inner viscous core. In figure~\ref{cp0_Long_Hall}, typical distributions of $C_{p_0}$ versus $\phi$ or both the generalized Hall-type flows studied by Mayer and Powell and for Long's infinite-domain conical solutions are shown, all for flows of similar strength as measured by the vortex-axis value of axial velocity excess. For the generalized Hall-type flows all the $C_{p_0}$ distributions are asymptotically constant in the far field and decrease monotonically as $\phi\ra 0$, but for Long's solutions, $C_{p_0}$ achieves a maximum on the vortex axis, as explained previously. As is clear from the similarity equations, the value of the pressure (and not just its derivative) plays a role for nonzero axial-flow exponents $m$ ($n\ne\half$). For the Hall-type solutions one can set $p$ independently of the velocities at the finite outer boundary $\phi_e$, but for Long-type solutions one can only specify that $p$ vanish as $\phi\ra\infty$ and then take whatever total-pressure distributions one gets from satisfying the similarity equations for the given on-axis value of axial velocity; it is clear the figure (and from analogous plots of other Long-type flows for all admissible values of flow force, not shown here) that at least for $n=1$ what one gets is a nonphysical $C_{p_0}$ profile.

%%% Thesis Fig 2.2, p39

%%%%%%%%%%%%%%%%%%%%%%%%%%%%%%
\section{Far-Field Asymptotics}
\label{sect:asymp}

In order to understand the far-field behavior of the generalized Long-type flows and - as it turns out - to permit application of the spectral numerical scheme described in section~\S\ref{sect:numerics} to solve the similarity equations, the far-field asymptotic form of the stream function $f$ and circulation $g$ for general values of $n$ is required. For $n=1$, following Long and including our slightly different but equivalent definition of the independent variable $\phi$ one finds that
\abceqnbeg
\begin{eqnarray}
f \sim \frac{\phi}{\sqrt{2}}+a_0+a_1\phi^{-1}+...,\\*
g \sim 1 + {\rm exponentially\ decaying\ terms}.
\label{asymp_long}
\end{eqnarray}
\abceqnend
(A detailed proof that the non-constant terms in the far-field expression for $g$ do in fact vanish exponentially fast for the $n=1$ case can be found in \cite{LeeThesis}.  Accordingly, all velocity components decay as ${\cal O}(\phi^{-1})$, and the pressure as ${\cal O}(\phi^{-2})$. The coefficient $a_1$ is a continuous function of axis value of axial velocity $w_{axis}$, but begins to drop with ever-increasing rapidity as this parameter becomes slightly less than zero, and it appears there are no solution below a limiting lower-bound values of $w_{axis}$ which lies somewhere between -0.05 and -0.10, and coincides with a negative divergence of the $a_1$ coefficient.

In the general case, perhaps the simplest way to find the leading-order power of $\phi$ in the asymptotic series for $f$ is by requiring the radial velocity $u$ to decay at infinity (a property of the $n=1$ solutions, which we also require in the general case), i.e.~that
\begin{equation}
u=nf^\p-\frac{f}{\phi}\ra 0\ {\rm as}\ \phi\ra\infty,
\end{equation}
and $f\sim\phi^\frac{1}{n}$ is the largest power for which this is true. (For a more detailed analysis as to why this value is not merely an upper bound on the leading-order asymptotic-series-term exponent but in fact the unique proper choice, based on analysis of the inviscid terms in the scaled self-similar momentum equations, cf.~\cite{LeeThesis}). The relation between the leading-order terms in the $f$ and $g$-expansions can be found by integrating the inviscid form of the $\theta$-momentum equation~(\ref{sim_cooked2}) to yield $g\propto f^{1-n}\sim\phi^{\frac{1}{n}-1}$. With some further work (we again refer the interested reader to the thesis of Lee for the details here), it can be shown that the proper generalized form of the far-field asymptotic series for $f$ and $g$ are ones in which each successive term is a power $\phi^{-\frac{1}{n}}$ smaller than the preceding one:
\abceqnbeg
\begin{eqnarray}
&&f\sim f_{as}=a_{-1}\phi^{\frac{1}{n}}+a_0+a_1\phi^{-\frac{1}{n}}+a_2\phi^{-\frac{2}{n}}+...,\\*
&&g\sim g_{as}=b_{-1}\phi^{(\frac{1}{n}-1)}+b_0\phi^{-1}+b_1\phi^{(-\frac{1}{n}-1)}+b_2\phi^{(-\frac{2}{n}-1)}+... .
\label{asymp_gen}
\end{eqnarray}
\abceqnend
Before developing expressions relating the undetermined coefficients in these series, recall that for $n=\half$ and $n=1$ the $z$-momentum equation~(\ref{sim_cooked3}) is integrable. This raises the question: which form of the $z$-momentum equation should one use in the asymptotic analysis? This question turns out not have the trivial answer of ``either''. For $n=\half$ the above generalized far-field asymptotic series take the form
\begin{equation}
f\sim a_{-1}\phi^2+1+\ldots,\qquad g\sim\phi+\ldots .
\end{equation}
The corresponding swirl and axial velocities have the form
\begin{equation}
v=\frac{g}{\phi}\sim 1,\qquad w=\frac{f^\p}{\phi}\sim 2a_{-1},
\end{equation}
and thus do not decay at infinity. (For $n>\half$ they in fact diverge.) Thus, it appears that solutions with vanishing far-field velocities only exist if $n>\half$, that is, if the external axial pressure gradient is adverse ($m<0$), and from here on we shall consider only such cases in the context of Long-type solutions which are potentially physically realizable, although we can find and examine solutions with nondecaying velocities using our hybrid numerical/asymptotic solution procedure, as well.

For $n=1$ one should use the integrated third-order form~(\ref{int_one}) of the equation. This is not merely for the convenience of using a lower-order form of the equation; as it happens, if one tries to use the fourth-order form of the equation when $n=1$, one cannot solve for the leading-order coefficient in the $f$-series, since the relevant terms in the asymptotic hierarchy cancel identically. When one takes $n=1$, substitutes the expansions for $f$ and $g$ into equations~(\ref{sim_cooked2}) and~(\ref{int_one}) and collects terms at various orders in $\phi$, one obtains the following expression hierarchy for the coefficients, which can be solved one line at a time:
\begin{eqnarray}
1-2a_{-1}^2=0;&&\nonumber\\*
a_{-1}b_0 = 0,&\quad&3(a_0-1)a_{-1}-2b_0=0;\nonumber\\*
2a_{-1}b_1-b_0(3-a_0) = 0,&\quad&b_0^2+2b_1=0;\\*
3a_{-1}b_2-2b_1(4-a_0)+a_1 b_0=0,&\quad&5\bigl[a_1(3-a_0)-a_{-1}a_2\bigr]-2(b_0 b_1+b_2)=0;\nonumber\\*
4a_{-1}b_3-3b_2(5-a_0)+2a_1 b_1+a_2 b_0=0,&\quad&6\bigl[2a_2(4-a_0)-2a_{-1}a_3-a_1^2\bigr]-2(b_0 b_2+b_3)-b_1^2=0; ...\nonumber
\end{eqnarray}
from which we see immediately that $a_{-1}=1/\sqrt{2}$, $a_0=1$, $a_1$ is indeterminate, $a_2=2\sqrt{2} a_1$, $a_3=a_1(12-a_1/\sqrt{2})$, and similarly, all higher-order $a$-series terms are dependent on the a priori unknown $a_1$ term~-- in terms of our soon-to-be-described solution procedure, $a_1$ is the eigenvalue which must be found by matching to the details of the inner core solution. All the b-coefficients are zero, another indication of the aforementioned exponentially-fast asymptotic approach of circulation to its value at infinity.

\begin{figure}
\begin{centering}
\includegraphics{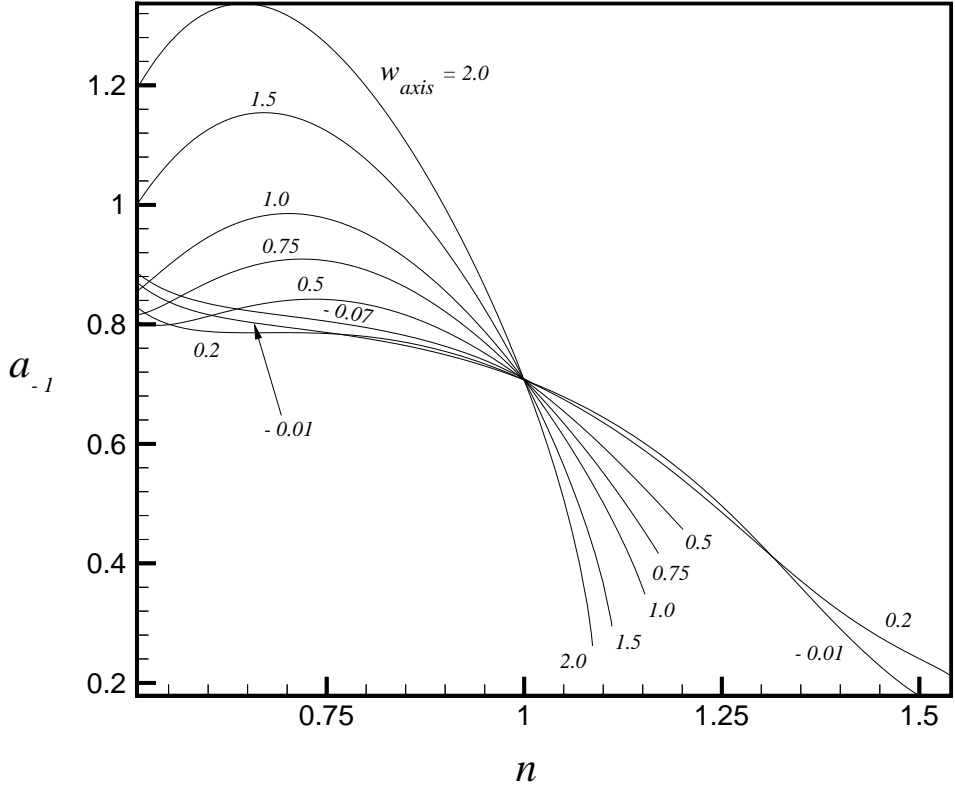}
\vspace{-0.3in}
\caption{Behavior of the coefficient $a_{-1}$ in the asymptotic series $f_{as}$ with respect to $n$ for various values of $w(0)$. }
\label{am1_vs_n}
\end{centering}
\end{figure}

In the more-general nonintegrable case one encounters a similar scenario, but now $a_{-1}$ is the undetermined coefficient and all higher-order coefficients are functions of it. After some rearrangement the first few $f$-series coefficients are:
\begin{eqnarray}
a_0&=&\frac{2n^3(1-n)(1-3n)-a_{-1}^2(1-2n)^2(1-4n)}{a_{-1}^2 n(1-2n)(1-4n)-2n^4(1-n)},\nonumber \\*
a_1&=&\frac{b_0 n^2[n(a_0-3)-a_{-1}b_0]}{2(1-n)(4a_{-1}^2+n^2)},\label{fcoeff_gen} \\*
a_2&=&\frac{N_{21}-N_{22}}{3a_{-1}^2(3-2n)(1+4n)+6n^3(1-n)}, ... ,\nonumber \\*
{\rm where}\;N_{21}&=&a_{-1}a_1(1+2n)(1+4n)\left(2-a_0+\frac{1}{n}\right),\nonumber \\*
N_{22}&=&2n^3\left[3a_{-1}b_0 b_1+(1+n)\left(3-a_0+\frac{1}{n}\right)b_1+a_1 b_0(1-2n)\right].\nonumber
\end{eqnarray}
Note that all higher-order terms in the two generalized asymptotic series can be expressed strictly in terms of $a_{-1}$, but as this quickly leads to extremely unwieldy expressions, we have instead chosen to write each higher-order coefficient in terms of convenient combinations of lower-order terms in order to keep the notation reasonable. This kind of recursive cascade of coefficients also proves convenient from the standpoint of numerical implementation~-- the dependencies among the terms as written merely imply that they must be calculated in a particular order, i.e.~provided with a suitable value (or initial guess) for the leading-order term $a_{-1}$, we calculate in order $a_0$, $b_0$, $a_1$, $b_1$, and so forth when we update the coefficients after each iteration.
%
%As a consistency check, if one sets $a_{-1}=1/\sqrt{2}$, $n=1$ and all the b-coefficients equal to zero in the above expressions, one can quickly verify that they reduce to same form as for the special case of Long.
%
The first few coefficients in the generalized far-field $g$-series are
\begin{eqnarray}
b_{-1}&=&1,\nonumber \\*
b_0&=&\frac{(1-n)}{a_{-1}}\left(a_0+\frac{1}{n}-3\right),\nonumber \\*
b_1&=&\frac{1}{2a_{-1}}\biggl[n(3-a_0)b_0+2a_1(1-n)\biggr],\label{gcoeff_gen} \\*
b_2&=&\frac{1}{3a_{-1}}\left[a_1 b_0(1-2n)-(1+n)\left(a_0+\frac{1}{n}-3\right)b_1+3a_2(1-n)\right], ... .\nonumber
\end{eqnarray}
\abceqnend
This difference in indeterminate coefficient seems strikingly at odds with the integrable $n=1$ case, especially since as $n$ varies we expect a continuum of solutions. Reassuringly, in the general case the dependence of $a_{-1}$ on the details of the viscous core becomes less and less sensitive as $n\ra 1$ until at $n=1$, numerical solutions of the full (nonintegrated) equations for any value of vortex strength (as measured by either flow force or $w_{axis}$) all yield a single numerical value for $a_{-1}=1/\sqrt{2}$, which is exactly the unique value found from doing the asymptotics using the integrated form of the $z$-momentum equation for this special case. This behavior is illustrated in figure~\ref{am1_vs_n}. Thus, even though the undetermined coefficients appear in different places depending on whether $n=1$ or not, the two forms agree for the unique value of $n$ for which both are applicable.
These formulae show that when $n=1$, except for $a_{-1}$ not being determinable {\it a priori}, the coefficients given by the general-$n$ asymptotics agree with those for the $n=1$ integrated-equation results if one sets $a_{-1}=1/\sqrt{2}$. Note also that the factor $(1-n)$ in the numerator of $b_0$ and the denominator of $a_1$ makes $a_1$ indeterminate when $n=1$. Thus the above are only strictly valid for $n\ne 1$. However, as we shall see, the above coefficients are consistent with those for the $n=1$ case, in that the (in general indeterminate) coefficient $a_{-1}$ converges to $a_{-1}=1/\sqrt{2}$ for all inner-core solutions (i.e. vortex strengths) as $n\ra 1$.

%%% Thesis Fig 3.1, p51

\begin{figure}
\begin{centering}
\includegraphics[scale=0.7]{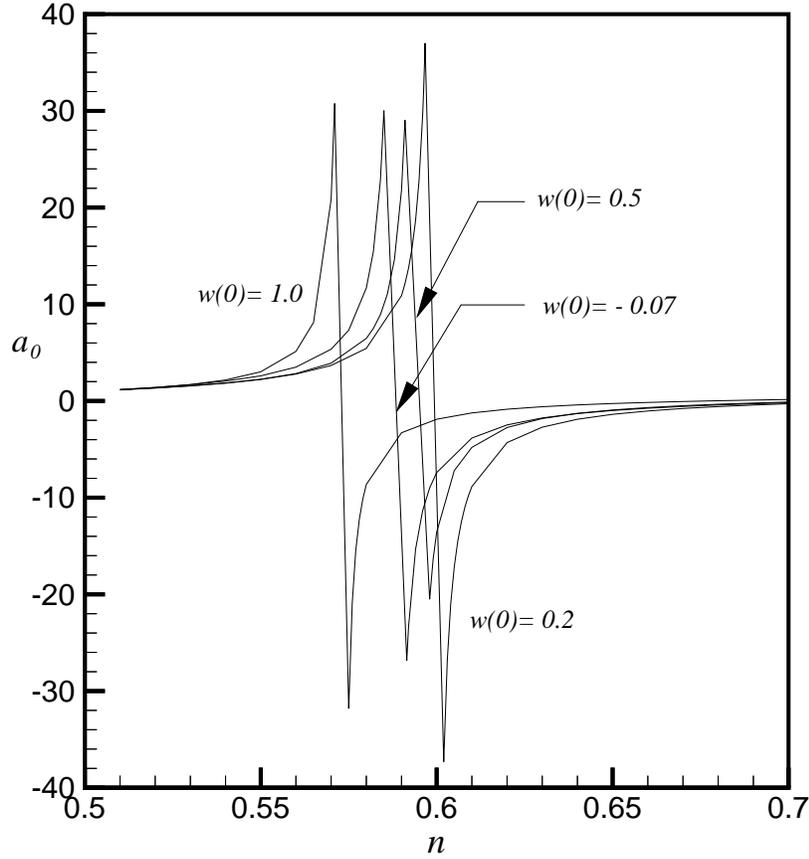}
\vspace{-0.2in}
\caption{The singular behavior of $a_0$ for $w_{axis}=1.0,\ 0.5,\ 0.2$, and $-0.07$.}
\label{coeff_si}
\end{centering}
\end{figure}

Lastly, before describing the numerical method used by us to solve the generalized equations, we note that there is one additional ``asymptotic surprise'' in store. Inspection of the expression for $a_0$ in~(\ref{fcoeff_gen}) indicates that it is possible for the denominator to vanish for values of the growth-rate parameter $\half<n<1$ if the value of the leading-order coefficient $a_{-1}=$ (which depends continuously on the vortex strength, as measured by the value of $w(0)$) happens to take on the value
\begin{equation}
a_{-1}=\pm\sqrt{\frac{2n^3(1-n)}{(1-2n)(1-4n)}},
\end{equation}
%
% Sample values of x = 
%  n	x		sqrt(x)
% ----  ------	------
% 0.50	oo
% .51	2.49998057684762515344
% .52	1.76766009804549874107
% .53	1.44308892822372688381
% .54	1.24942607514065405403
% .55	1.11705528063744454956
% .56	1.01911198958300356019
% .57	.94274124186180741052
% .58	.88091636173000916945
% .59	.82942904017145346276
% .60	.78558440484957257255
% .61	.74756587325030083162
% .62	.71409805771759107623
% .63	.68425537263492281987
% .64	.65734731649676223918
% .65	.63284657830051247473
% .66	.61034226703287497971
% .67	.58950850130362301457
% .68	.57008272666822491278
% .69	.55185038343175943135
% .70	.53463383107818133409
% .71	.51828419244733554720
% .72	.50267524153533881657
% .73	.48769874711217231984
% .74	.47326086922556364506
% .75	.45927932677184589341
% .76	.44568113512706009123
% .77	.43240076756311821927
% .78	.41937863161787575832
% .79	.40655977726580559500
% .80	.39389277113386474474
% .81	.38132868229401806419
% .82	.36882013160032928240
% .83	.35632035868671474087
% .84	.34378225852444957945
% .85	.33115733205144133033
% .86	.31839448100954988020
% .87	.30543855231699151711
% .88	.29222849576031396940
% .89	.27869492887637955083
% .90	.26475678243654844309
% .91	.25031648481884305226
% .92	.23525274104191169814
% .93	.21940916418178476895
% .94	.20257531659429939639
% .95	.18445275129506054703
% .96	.16458822414513882611
% .97	.14222405470801573399
% .98	.11588896039551183845
% .99	.08179163624426776712
% 1.00	0
%
where the argument of the square root descends from a positive divergence as $n$ increases from 0.5, and is real and positive for all $n$ lying strictly between one-half and unity. For $n=1$ it might appear at first glance that $a_1$ similarly blows up, but substitution of the expression for $b_0$ reveals that the zero denominator is canceled by an equal-order vanishing numerator, leaving a finite result, albeit an indeterminate one.  Note also that for the above values of $a_{-1}$, the numerator in the expression for $a_0$ takes the value $-2n^4(1-n)$, so the zero denominator is not balanced by a vanishing numerator unless $n=0$ (outside the range of consideration) or $n=1$, the conical Long's case. Some of this singular behavior of $a_0$, here plotted for constant vortex strength and with varying $n$, is illustrated in figure~\ref{coeff_si}, here by plotting the evolution of $a_0$ for various fixed values of $w_{axis}$ as the vortex-growth-rate parameter $n$ is varied. (The line segments connecting the positive and negative-divergence portions of each trend line are artifacts of the plotting software, but useful ones in terms of connecting the two distinct asymptotic branches of each curve). We were in fact first alerted to this possibility by way of seemingly curious behavior of the numerical iteration procedure for values of vortex strength which happened to correspond to values of $a_{-1}$ close to the above ``blowup value'' and which thus resulted in near-singular behavior in $a_0$ and the other terms depending on it. Interestingly, the occurrence of such ``coefficients crises'' appears to be closely correlated with the appearance of near-axis overshoots of the total pressure coefficient of the solutions, that is, with incipient nonphysical behavior. It is possible that the coefficient singularities define separatrices dividing physically possible from nonphysical flow regimes, with the "physical" portion of the solution space lying on the lower-$n$ side of each such divergence point.

%%% Thesis Fig 3.3, p58

%%%%%%%%%%%%%%%%%%%%%%%%%%%%%%
\subsection{Conservation of Axial and Angular Momentum Flux In a Constant-z Plane}
\label{sect:momentum}

For any flow field of the type under consideration here, since there is no axial external impulse or torque acting on the flow, one might at first blush expect that the axial and angular momentum fluxes through any constant-$z$ plane should be constant for the flow to be be physically plausible. This is indeed the case for Long's conical solutions, but as it turns out, does not hold for the general nonconical similarity solutions. However, this appears to be a case in which a superficially reasonable-seeming criterion indeed proves not so, and the fact that the above properties hold for one special value of the core growth rate parameter $n$ (which is also exceptional in numerous other ways, as we have detailed) turns out to be a quirk of the asymptotic solution properties which obtain in that case, rather than anything having a bearing on physicality of the solutions. Because the analysis involved here illustrates some further important aspects of the flows and the similarity solutions, we shall devote a page or two to it.

To obtain the axial momentum flux through a constant-$z$ plane, we integrate the $z$-momentum equation~(\ref{zmomd}) with respect to the radial and polar $(\phi,\theta)$ coordinates of the similarity formulation:
\begin{equation}
\int_0^\infty [uw^\p+w(mw-n\phi w^\p)]\phi d\phi = \int_0^\infty \left[\left(w^\pp+\frac{w^\p}{\phi}\right)+n\phi p^\p-2mp\right]\phi d\phi .
\end{equation}
On using the continuity and r-momentum equations equations~(\ref{massd},b), integration by parts and some rearrangement, we obtain:
\begin{equation}
2(m+n)\int_0^\infty (p+w^2)\phi d\phi = \bigl[\phi w^\p+n\phi^2 p-\phi w(u-n\phi w)\bigr]\biggr|_0^\infty .
\end{equation}
Since the right-hand side of this last expression is zero at $\phi=0$, we need only consider the limit at infinity, for which we can make use of the asymptotic series for stream function and circulation for general $n$ which we derived in the preceding section. Using that $u=nf^\p-f/\phi$, $v=g/\phi$ and $w=f^\p/\phi$, we have:
\begin{equation}
2(m+n)\int_0^\infty (p+w^2)\phi d\phi \sim \left[\frac{a_{-1}^2}{n}+\frac{n^2}{2(1-2n)}\right]\phi^{(\frac{2}{n}-2)} + {\cal O}(\phi^{(\frac{1}{n}-2)}).
\end{equation}
The first thing to note about this expression is that is shows why the far-field asymptotic constant $a_{-1}$ tends to a fixed value as $n\ra 1$ (and thus $m\ra -1$: For $(m+n)\ra 0$, the entire left-hand-side of the above expression vanishes, hence $a_{-1}$ {\em must} tend to $1/\sqrt{2}$ (the value which makes the leading-order term of the right-hand side vanish), independently of whatever core-flow solution details are contained in the terms appearing in the left-hand-side integrand.
The above integral is~-- up to a constant involving the common powers of $\eps$ and $z$ which we eliminated from both sides of the scaled $z$-momentum equation during the similarity formulation~-- equivalent to the axial derivative of flow force. It is clear that the leading term on the right-hand side of the above asymptotic expression only decays at infinity for $n>1$. For $n=1$ the power of $\phi$ of this term is precisely zero, but the coefficient vanishes since $m+n$ does. For $\half<n<1$, on the other hand, this term diverges unless
\begin{equation}
a_{-1}=n\sqrt{\frac{n}{2(2n-1)}}.
\end{equation}
The curve described by this equation in fact passes through the physically-plausible region of the solution space based on the $C_{p_0}$ criterion. Thus, if we insist on constant axial momentum flux the best we can say is that for $n<1$ there are at least {\it some} points in the physical region (based on total pressure coefficient) which also satisfy the requirement of constant axial momentum flux. But it is in fact not reasonable to impose such a requirement, and further it makes no sense that there be nonphysicality related to conservation of momentum, because the similarity formulation by construction conserves momentum, at least up to the order of the neglected terms in the small parameter. Regarding the artificiality of the constant-axial-momentum-flux criterion, we quote \cite{Morton} (emphasis his):
\begin{quotation}
We note that the pressure $p_\infty-p_0$ on the axis of a rotating core embedded in an externally still environment at uniform pressure $p_\infty$ is constant only in a cylindrical field $v=v(r)$ of azimuthal velocity, and that there will always be axial pressure gradients (with corresponding axial velocities) in rotating cores exhibiting axial development of the azimuthal velocity field $v=v(r,z)$. Thus {\em axial pressure gradients are the rule in all swirling flows suffering progressive lateral spread by either viscous or turbulent diffusion} ...
In strongly rotating cores the pressure thrust is of comparable importance to the momentum flux, and the axial pressure gradients normally formed result in a progressive transfer of force between the momentum flux and pressure fields ... in the absence of body forces the axial changes in flow force are due to entrainment of ambient axial momentum and to outer viscous stresses.
\end{quotation}

In the case of  variation of angular momentum flux in the axial direction, we obtain an analogous result~-- the flux is constant for the special conical case due to the exponential approach of the circulation to a constant in the far field and corresponding vanishing of the higher-order {\em b}-coefficients in~(\ref{asymp_gen}), nonconstant in the general case, but satisfying the momentum conservation equations up to the neglected terms in the similarity equations. As to the physical mechanisms for the nonconstant angular momentum flux, again quoting Morton:
\begin{quotation}
...the axial rate of change in the flux of angular momentum (due primarily to convection, and in negligible proportion ... to viscous diffusion) is the result of an effective torque due to entrainment of azimuthal momentum at infinity and to the moment of viscous stress at infinity.
\end{quotation}

%%%%%%%%%%%%%%%%%%%%%%%%%%%%%%
\section{Numerical Method}
\label{sect:numerics}

We desire our numerical solutions to be of sufficient accuracy to serve as base-flow profiles for a subsequent numerical linear stability analysis, thus we need the basic-flow solutions to be accurate to roughly the level of round-off error in the latter~-- typically that means to the level of IEEE 64-bit floating-point arithmetic. (In other words, a standard kind of finite-difference approximation scheme simply will not do). To that end, the general similarity equations~(\ref{sim_cooked2},c) are discretized using a Chebyshev polynomial basis and a standard interior-plus-endpoints meshing~(cf. \cite{BoydBook}) of the Chebyshev interval $x\in[-1,1]$, along with a suitable mapping transformation to the semi-infinite ``physical'' (that is, $\phi$) space:
\begin{equation}
x_j=-{\rm cos}\left[\pi\frac{j-1}{N}\right],\qquad{\rm and}\qquad\phi_j=A\frac{1+x_j}{1-x_j},\qquad{\rm for}\qquad j=1,\ldots,N,
\end{equation}
where $N$ is the number of spectral collocation points and $A$ is a computational parameter (typically order of unity) we dub the ``half-points radius'' because it coincides with the physical location inside which precisely half the collocation points are located. The parameter $A$ can be tuned to give a desirable level of resolution of the immediate near-axis region, according the observed behavior of the solutions for some intermediate level of numerical resolution. We use the rapidity of convergence of key numerical solution parameters to guide the choice of $A$~-- the speed of convergence of the asymptotic-series coefficients $a_{\pm 1}$ with increasing $N$, for example, appears to be a good guide to the choice of an optimal value of $A$ for a given set of flow parameters.

This nonlinear one-parameter mapping is designed to spread the outermost collocation points (which have a cosine-type clustering in computational space, i.e.~are most densely clustered around the endpoints $x=\pm 1$), far apart as $\phi\ra\infty$, while still maintaining good spatial resolution near the rotational-symmetry axis of the flow, that is the origin of our radially semi-infinite coordinate system. Our preliminary numerical experiments showed that this stretched-grid method will only converge if the dependent variables being solved for decay at infinity. A little analysis shows that this is a direct consequence of the above nonlinear mapping between $x$ and $\phi$-space being singular at $x=1$, as a result of which the dependent variables must thus decay as $\phi\ra\infty$ sufficiently rapidly to cancel out this singularity in the mapping function. The generalized asymptotic series for $f$ and $g$ thus prove not only useful but in fact necessary for our full-field computations. We rewrite equations~(\ref{sim_cooked2},c) (substituting equation~(\ref{int_half}) for~(\ref{sim_cooked3}) in the special case $n=1$) in terms of modified functions $F$ and $G$, defined as
\begin{equation}
F(\phi):=f(\phi)-f_{as}(\phi)s(\phi),\qquad G(\phi):=g(\phi)-g_{as}(\phi)s(\phi),
\end{equation}
where $s(\phi)$ is a suitable interpolation function, designed to transition smoothly from zero at $\phi=0$ to unity as $\phi\ra\infty$, with exponential-asymptotic behavior in both limits. This gives the desired decay of the dependent variables at infinity, while still allowing one to use the original inner boundary conditions on $f$ and $g$; that is, the boundary conditions in term of the modified dependent variables are
\begin{equation}
\phi=0:F=F^\p=F^\pp-w_{axis}=G=0;\qquad F,G\ra 0 \;\;{\rm as}\;\; \phi\ra\infty.
\label{bc_mod}
\end{equation}
We find the following interpolation function to work well in practice:
\begin{equation}
s(\phi)=\frac{1+{\rm tanh}[z(x)]}{2}, \;\;{\rm where}\;\; z=\frac{Cx}{\sqrt{1-x^2}} \;\;{\rm maps}\;\; [-1,1]\;{\rm to}\;(-\infty,\infty).
\end{equation}
Here, $C$ is an order-unity constant which controls how quickly the function $s$ transitions from near-zero values to near-unity values. We want $s$ to approach unity exponentially fast as $\phi\ra\infty$, but we do not want the transition region about the inflection point to be too narrow, in order to maintain good numerical resolution there. After transforming the original system of equations for $f$ and $g$ into the modified equations for $F$ and $G$, we set up a Newton-type iterative scheme by linearizing the variables we wish to solve for, i.e.~the modified stream function and circulation $F$ and $G$ and the asymptotic coefficient $a_{\pm 1}$ (the $\pm$ sign on the subscript depending on whether $n=1$ or not, respectively) about their values at the hypothetical current ($k$th) iteration:
\begin{equation}
F^{(k+1)}=F^{(k)}+\delta F^{(k)},\qquad G^{(k+1)}=G^{(k)}+\delta G^{(k)},\qquad a_{\pm 1}^{(k+1)}=a_{\pm 1}^{(k)}+\delta a_{\pm 1}^{(k)},
\end{equation}
where an iteration cycle begins with some initial guesses for the functions being solved for $F^{(0)}$ and $G^{(0)}$ and for the numerical parameter $a_{\pm 1}^{(0)}$. Substitution into the modified equations and neglect of quadratic $\delta$-terms leads to system of linear differential equations for the functions $\delta F$ and $\delta G$ (where we drop the iteration superscripts to simplify the notation), with $\delta a_{\pm 1}$ appearing as an eigenvalue and the $k$th-iteration residuals (which are just the modified equations written entirely in terms of the known current-iteration values $F^{(k)}$, $G^{(k)}$ and $a_{\pm 1}^{(k)}$) appearing as forcing terms. Next, we expand each of the two unknown functions in an $N$-term-truncated Chebyshev expansion:
\begin{equation}
\delta F(\phi)=\sum_{j=0}^{N-1}c_j T_j(x(\phi)), \qquad \delta G(\phi)=\sum_{j=0}^{N-1}d_j T_j(x(\phi)),
\end{equation}
where $T_j$ is the $j$th Chebyshev polynomial, which one can define conveniently via the trigonometric identity $T_j(\cos y) = \cos(jy)$, although in practice we use the standard 3-term recurrence
\begin{equation}
T_{j+1}(x)=2xT_j(x)-T_{j-1}(x) \;\;{\rm for}\;\; j > 1, \;{\rm with}\; T_0=1 \;{\rm and}\; T_1=x,
\end{equation}
in order  to generate the numerical values of the $T_j$ and the needed derivatives thereof at the collocation points. Note that if one had reason to believe that there were a multiplicity of solutions at given flow parameters (with solutions as parameterized here, i.e.~with the non-uniqueness resulting from use of flow force as the key parameter removed) it would be worth considering a more-general eigensystem approach, but ones such as are used to investigate multiple eigenmodes in linear stability analyses seem inapplicable due to the nonlinearity of the equations in question. However, as we have found no evidence of multimodality in our $w_{axis}$-based parameterization (e.g.~via any of the quite different kinds of initial guesses we tried early in our investigations at the same values the flow parameters $w_{axis}$ and $n$ converging to different solutions), this iterative single-value technique for finding both $a_{\pm 1}$ and the Chebyshev-expansion coefficients of the associated functions $F$ and $G$ proved sufficient for the task at hand, in addition to being computationally fast.

Requiring the expansions to satisfy the linearized equations at each of the $N-2$ interior collocation points, along with the five boundary conditions~(\ref{bc_mod}), also rewritten in linearized $\delta$-form, yields a total of $2\times(N-2)+5=2N+1$ linear-algebraic equations for the $2N+1$ unknowns (the $2N$ expansion coefficients $c_j$ and $d_j$ and the iterative correction to the asymptotic coefficient $\delta a_{\pm 1}$), hence a well-posed linear system, which is solved by standard numerical techniques~-- we use Gaussian elimination with full (row and column) pivoting. We perform the computation in 128-bit emulated floating-point arithmetic with roughly 34 decimal digits of precision, and stop the iteration when all terms in the discrete residuals vector are less than $10^{-25}$ in magnitude. Beginning with $N=60$, we repeat the entire iteration sequence (assuming it converges), each time with a larger $N$ (increasing by 20 up to 120, then proceeding in increments of 50 from $N=150$), using the converged numerical solution from the next-smaller $N$ as an initial guess, until the values of the asymptotic coefficient $a_{\pm 1}$ obtained using successive values of $N$ agree to at least 10 significant figures.

%%% Thesis Table 4.1, p70
\begin{table}
\begin{center}
\begin{tabular}{r|l|l|l|l|l|l}
$N$&$\qquad {\cal O}(1/\phi^2)$&RelErr&$\qquad {\cal O}(1/\phi^3)$&RelErr&$\qquad {\cal O}(1/\phi^4)$&RelErr\\
\hline
    60& \ \ 1.61906864379741&3.8e0	& \ \ 1.66384839604259&3.9e0	& \ \ 1.76224789083692  &4.1e0\\
    80&$-$0.557871080988182 &2.7e-2	&$-$0.557983406848442 &2.7e-2	&$-$0.558338891261082   &2.6e-2\\
   100&$-$0.573348557424447 &1.3e-4	&$-$0.573349078388717 &1.3e-4	&$-$0.573327253499490   &1.6e-4\\
   120&$-$0.573421819792454 &4.8e-7	&$-$0.573421791543461 &5.2e-7	&$-$0.573422043738902   &8.5e-8\\
   150&$-$0.573422095791405 &6.1e-9	&$-$0.573422092143038 &3.1e-10	&$-$0.573422109942654   &1.3e-8\\
   200&$-$0.573422093404780 &1.9e-9	&$-$0.573422092321886 &1.1e-13	&$-$0.573422092266259   &9.7e-11\\
   250&$-$0.573422092765454 &7.7e-10&$-$0.573422092321968 &3.7e-14	&$-$0.573422092321832   &2.0e-13\\
   300&$-$0.573422092535824 &3.7e-10&$-$0.573422092321954 &1.3e-14	&$-$0.5734220923219465  &3.0e-16\\
   350&$-$0.573422092437391 &2.0e-10&$-$0.573422092321949 &4.1e-15	&$-$0.573422092321946679&7.6e-18
%%%400&$-$0.573422092389617 &		&$-$0.5734220923219479&			&$-$0.5734220923219466746737
\end{tabular}
\end{center}
\vspace{-0.2in}
\caption{Convergence of the asymptotic coefficient $a_1$ with increasing spectral resolution, with respect to the high-resolution reference value $(a_1)_{\rm ref}=-0.57342209232194667467\ldots$.}
\label{n1_converge}
\end{table}
Convergence histories of $a_1$ for three sets of numerical trials, all for the same flow parameters ($n=1$, $w_{axis}=1$, i.e.~Long's vortex with a central axial velocity of unity, and a collocation mesh with half-points radius $A=5$) are shown in Table~\ref{n1_converge}. The three trials differ only in the number of leading terms of the full asymptotic far-field series for $f$ used to construct $f_{as}$. The data show that in each case, the value of $a_1$ tends to the same constant as the numerical resolution is increased, and that retaining higher-order terms in the far-field asymptotic expansion (which has as its price an increase of the algebraic complexity of the modified equations and their discretization~-- we refer the reader to the source code listing in the appendix of \cite{LeeThesis}, especially the segment of the listing beginning on page 142, for the gory details) generally leads to faster convergence, especially at the higher values of $N$. (Note however the slightly anomalous behavior of the ${\cal O}(1/\phi^3)$ approximation in the range $N=150$ to 250, for which the respective relative errors are less than for the ${\cal O}(1/\phi^4)$ asymptotic series truncation, for reasons unclear.) In the table, the number of significant figures was computed by comparison with the reference value $a_1=-0.57342209232194667467\ldots$ yielded by a trio of ultra-high-resolution runs with $N=400$, 450 and 500 and asymptotic $f$-term up to ${\cal O}(1/\phi^4)$, which agreed to at least 20 significant figures. For the purpose of our general study of these solution families, however, 5 to 6 significant figures in the requisite asymptotic constant $a_{\pm 1}$ appears to be more than adequate, so for our general-$n$ runs we have retained terms up to ${\cal O}(\phi^{-\frac{2}{n}})$ in $f_{as}$ and ${\cal O}(\phi^{-\frac{2}{n}-1})$ in $g_{as}$, that is, the leading four terms in each of the generalized far-field asymptotic series~(\ref{asymp_gen}).

On a typical gigahertz-class workstation with reasonably efficient emulation of 128-bit floating-point arithmetic a full convergence cycle (i.e.~iterate at fixed $N$ until convergence of the  numerical residuals is achieved, increase $N$ and repeat until $a_{\pm 1}$ converges to at least 6 figures) for cases similar to those described above, with spectral resolutions ranging from $N=60$ to 300, takes under a minute. The higher-resolution runs are where most of the expense is incurred (since the work in solving the linear system at each iteration scales as ${\cal O}(N^3)$), but the extra work per iteration is partly offset by the fact that the higher-resolution runs, being bootstrapped from ever-higher-quality converged solutions for the next-smaller $N$, need just one or two iterations to converge. Low-resolution cases may take several tens of iterations, or even a few hundred in generalized Long's vortex cases where a good initial guess is not available or one is near the asymptotic-coefficients singularities mentioned above and the iterations need to be severely under-relaxed (e.g.~by updating as $F^{(k+1)}=F^{(k)}+\omega\delta F^{(k)}$, where the relaxation factor is in the range $0<\omega< 1$) in order to achieve convergence.

%%%%%%%%%%%%%%%%%%%%%%%%%%%%%%
\section{Discussion of Solutions of the Generalized Long-Type Flows}
\label{sect:solutions}

\subsection{The Solutions for $n=1$}
\label{sect:neq1}
\begin{figure}
\begin{centering}
\includegraphics[scale=0.9]{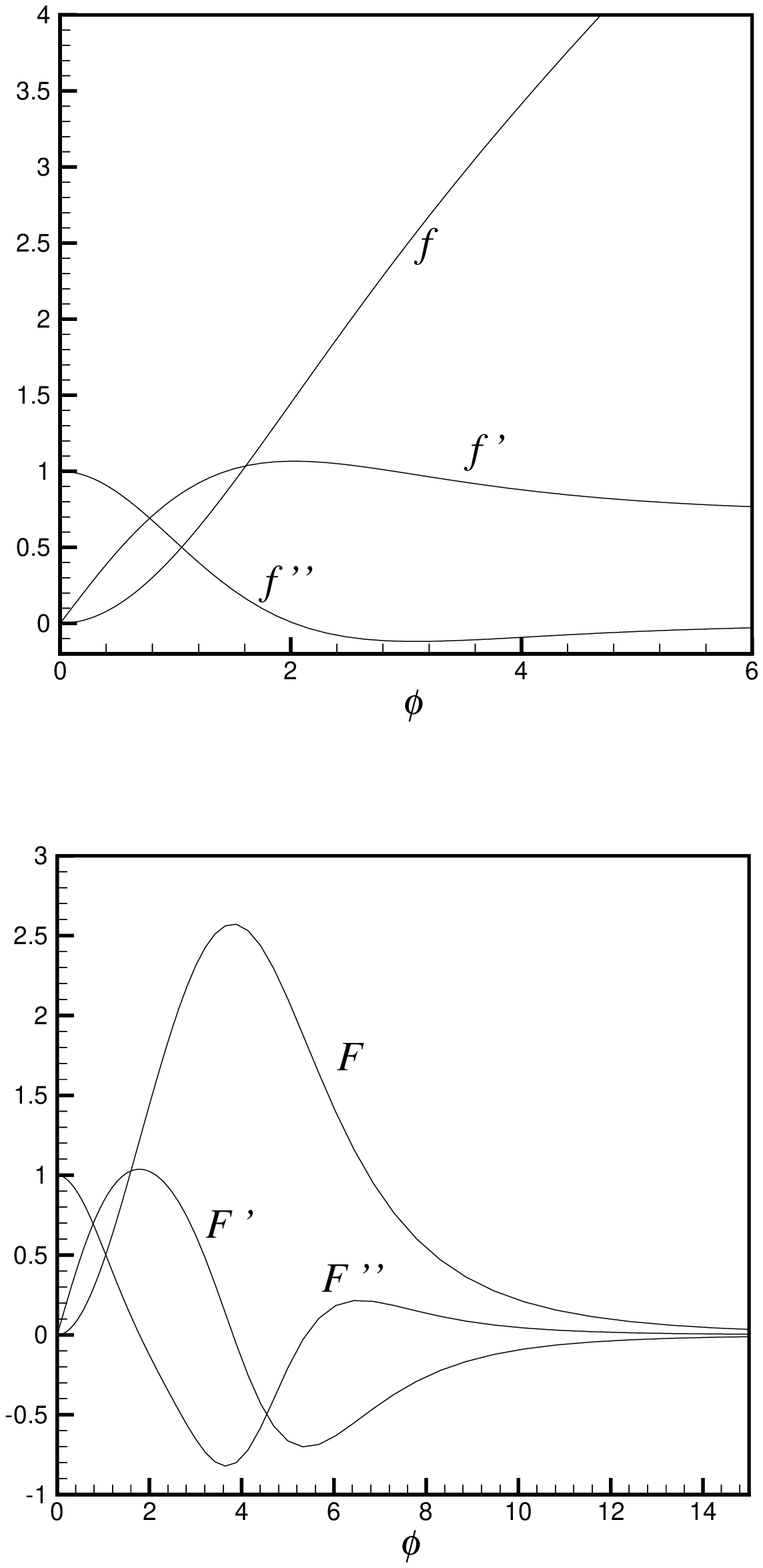}
\vspace{-0.2in}
\caption{Unmodified ($f$) and modified ($F$) stream functions and their first two derivatives for Long's flow ($n=1$) with $w_{axis}=1.0$.}
\label{n1_f_f_w}
\end{centering}
\end{figure}

\begin{figure}
\begin{centering}
\includegraphics[scale=0.9]{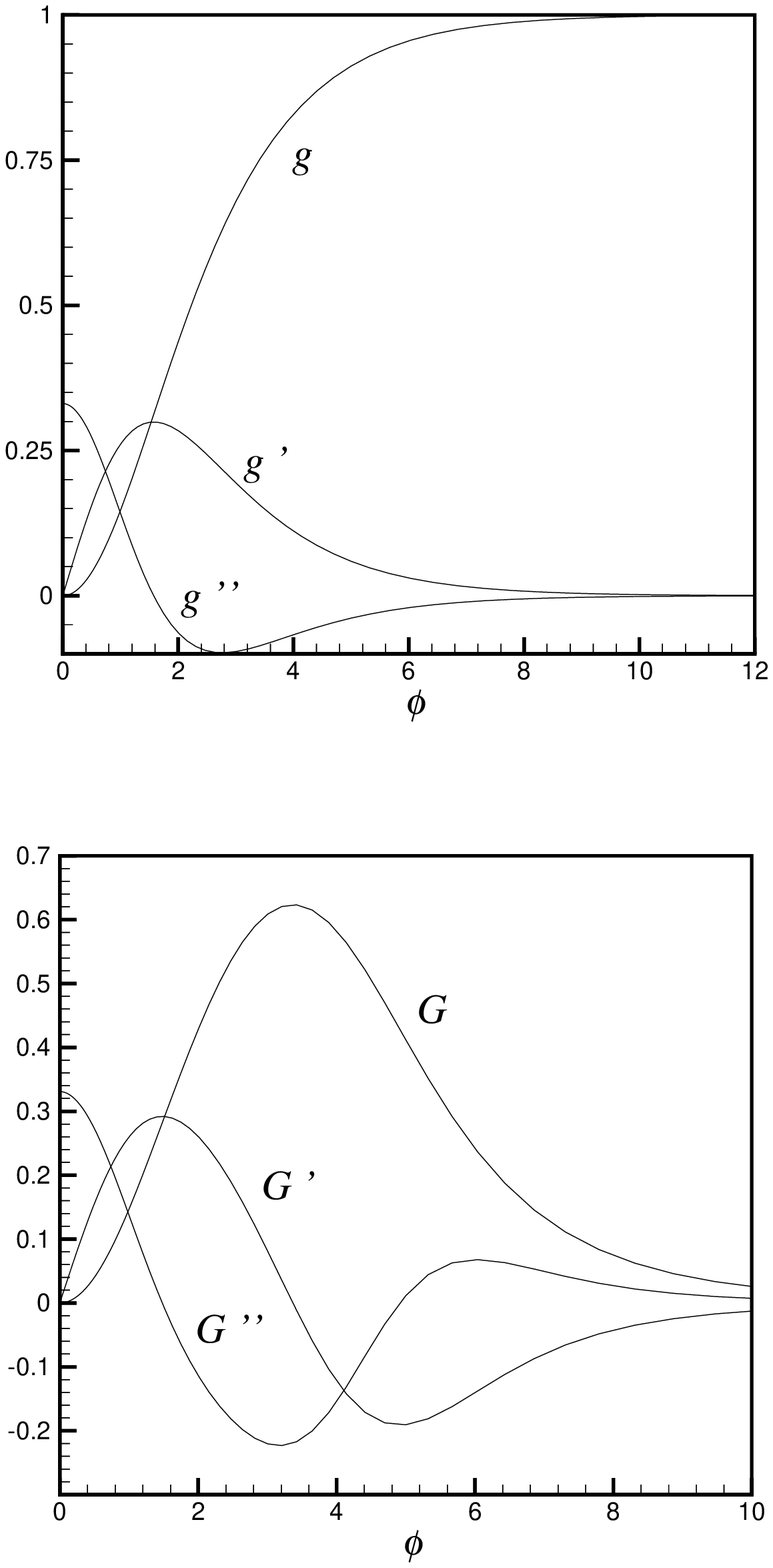}
\vspace{-0.2in}
\caption{Unmodified ($g$) and modified ($G$) circulations and their first two derivatives for Long's flow ($n=1$) with $w_{axis}=1.0$.}
\label{n1_g_g_w}
\end{centering}
\end{figure}

We first examine the solutions for Long's vortex ($n=1$), first for the special case of a vortex with a core axial velocity of unity. Unless otherwise noted, all of the plots represent solutions calculated with 200 basis functions for each of $F$  and $G$ and terms up to ${\cal O}(1/\phi^3)$ retained in the $f$-series, which imply solutions converged to well within "plotting accuracy". The modified and unmodified stream functions $F$ and their first two derivatives are plotted in figure~\ref{n1_f_f_w}.
As expected, $F$ and $F^\p$ are both zero at the axis, $F^\pp(0)$ is equal to the desired core axial velocity, and $F$ and all its derivatives vanish at infinity. The same qualitative behavior holds if we retain more or fewer terms in $f_{as}$, the only difference being that the modified functions $F$ decays more or less quickly in the far field. Analogous plots of the modified and unmodified circulation appear in figure~\ref{n1_g_g_w}.

%%% Thesis Fig 5.1,5.2, p72

%%% Thesis Fig 5.3,5.4, p73

\begin{figure}
\begin{centering}
\includegraphics[scale=0.7]{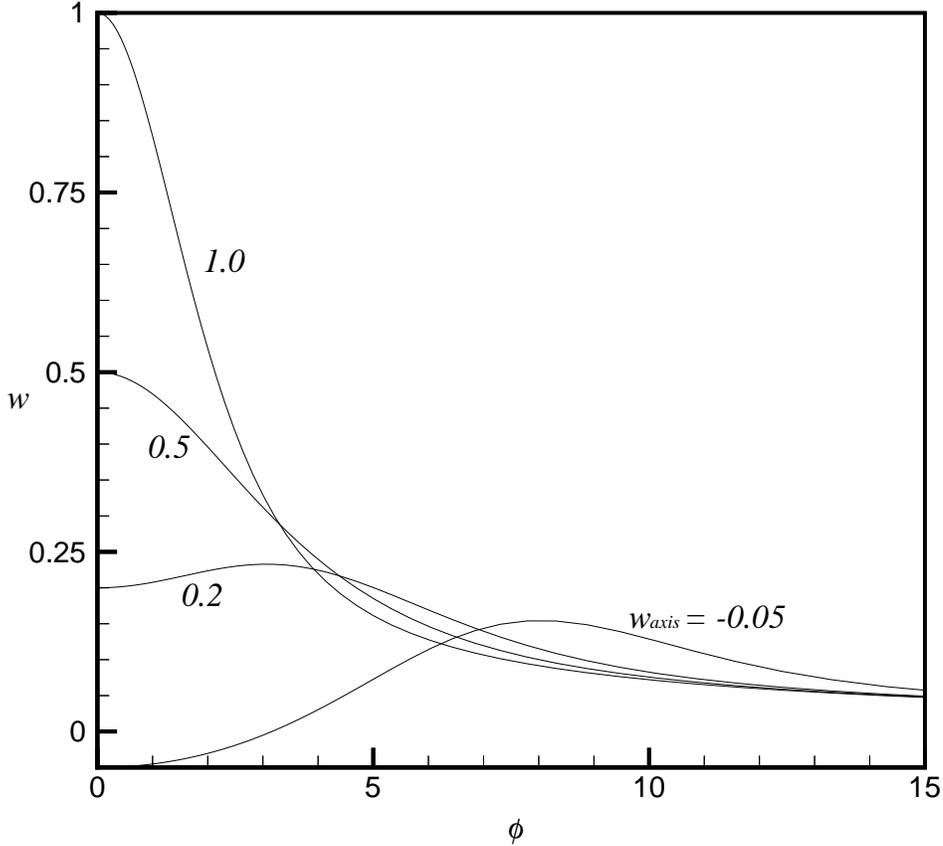}
\vspace{-0.2in}
\caption{Axial velocity distributions of Long's flow ($n=1$) for various values of $w_{axis}$.}
\label{n1_w_wva}
\end{centering}
\end{figure}

The distributions of the various velocity components for solutions with various values of core axial velocity ranging from $-0.05$ to 1 are shown in figures~\ref{n1_w_wva}-\ref{n1_u_wva}; by way of our boundary-condition parameterization we control the axis value of axial velocity but not whether it is a local maximum or minimum, i.e.~whether the axial flow is jet-like or wake-like. The axial velocity (fig.~\ref{n1_w_wva}) is clearly jet-like for $w_{axis}=1$ and 0.5, begins to show a core axial velocity deficit for $w_{axis}=0.2$, and for $w_{axis}=-0.05$ exhibits a region of reversed flow in the core.

%%% Thesis Fig 5.5, p75

The azimuthal velocity (fig.~\ref{n1_v_wva}) shows the expected transition from a solid rotation at the center to a potential-vortex-like decay in the far field, with a peak value that becomes smaller as the core axial velocity is reduced, but which remains positive even if $w_{axis}$ is less than zero. The swirl velocity distributions are much less sensitive to the flow parameters in the far field, in fact they all asymptote to the same curve, which is strictly determined by the normalization of nondimensionalized circulation in the far-field.

\begin{samepage}

%%% Thesis Fig 5.7, p76
\begin{figure}
\begin{centering}
\includegraphics[scale=0.6]{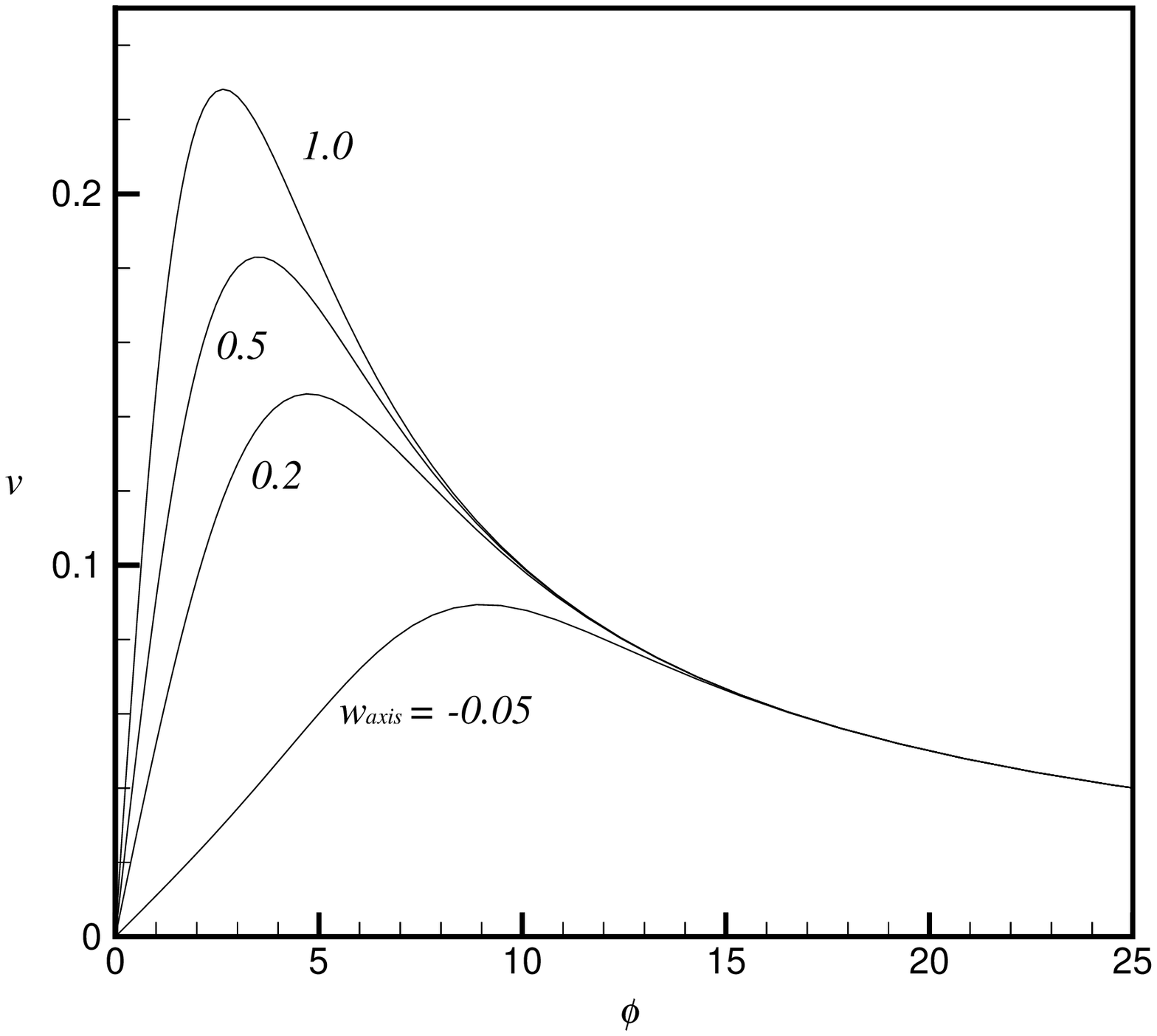}
\vspace{-0.2in}
\caption{Swirl velocity distributions of Long's flow ($n=1$) for various values of $w_{axis}$.}
\label{n1_v_wva}
%\end{centering}
%\end{figure}

%%% Thesis Fig 5.6, p76
%\begin{figure}
%\begin{centering}
\includegraphics[scale=0.6]{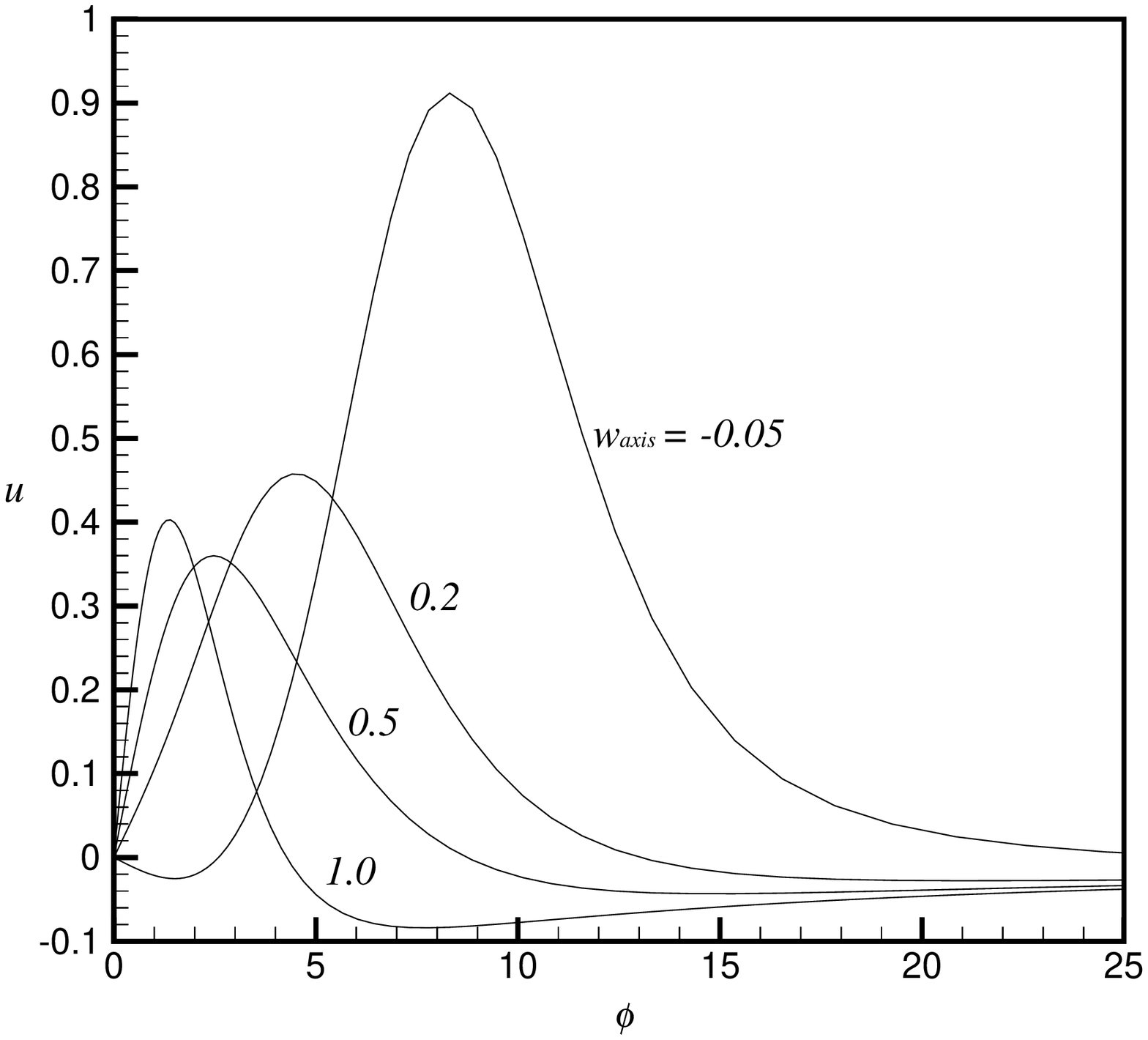}
\vspace{-0.2in}
\caption{Radial velocity distributions of Long's flow ($n=1$) for various values of $w_{axis}$. }
\label{n1_u_wva}
\end{centering}
\end{figure}

\end{samepage}

Owing to the requirement of conservation of mass the radial velocity curves in figure~\ref{n1_u_wva} show qualitatively different behavior for the different types of flows: if the axial velocity is jet-like in the core, the radial velocity is of positive sign near the axis, becomes negative in an intermediate region and tends to zero from below in the far field. If the axial velocity is wake-like in the core, the radial velocity is negative near the axis, becomes positive in an intermediate region and tends to zero from above in the far field.

As far as what happens when one tries to extend the parameter range further, this is quite different for the jet-like and wake-like solutions. For both, driving the core axis velocity further in the direction in which it is already tending (i.e.~making the flow more jet-like or wake-like, respectively) corresponds to increasing the flow force. Increasing $w_{axis}$ for the jet-like flows is not a problem, but making $w_{axis}$ more negative in the wake-like case becomes extremely difficult for $w_{axis}<-0.05$, and one must make successively smaller changes in this parameter value, and severely under-relax the iterations to have any chance at obtaining a converged solution. Figure~3 of \cite{FosterDuck}
%%%(see fig.~\ref{flowforce1})
provides a partial explanation of why this is so. It shows that whereas for the jet-like (Type~I) solutions, the slope of the solution curve in the flow force-$w_{axis}$ plane appears to tend to a constant positive value, for the wake-like (Type~II) solutions, $dM/dw_{axis}\rightarrow\oo$ as $M\rightarrow\oo$, meaning that small changes in $w_{axis}$ produce ever-larger changes in the flow force (and in certain of the solution curves). The other difficulty with increasingly wake-like solutions is related to the large-flow-force asymptotic properties of Type~II solutions described by \cite{FosterSmith}: as $M\rightarrow\oo$, these solutions take the form of a ring-jet, with an ever-narrower annular region of positive axial velocity centered at finite radius (and circulation rapidly transitioning from a value which is effectively zero elsewhere to unity in this interior layer). One can observe the beginnings of the development of this layer-like structure in figure~\ref{n1_w_wva}, for $w_{axis}=-0.05$. In conjunction with the flow force, the asymptotic coefficient $a_1$ also becomes increasingly sensitive to changes in the flow parameters as the flow becomes more wake-like. From a numerical standpoint (especially if one is using a Chebyshev-type spectral method, for which the set of basis functions tends to concentrate the spatial resolution near the endpoints of the computational domain), to achieve reasonable spatial resolution of the large-flow-force Type~II solutions, one would need to devise a modified mapping of the numerical to the physical domain allowing one to concentrate some desired fraction of the collocation mesh points inside the developing interior layer. This is not tremendously difficult to do, but we do not pursue it further, since our qualitative conclusions regarding the nonphysicality of the conical Long's solutions (see the discussion of the total-pressure profiles below) applies whether the the solution is jet-like or wake-like, and whether the flow force is order unity or asymptotically large.

%%% Thesis Fig 5.8, 79
\begin{figure}
\begin{centering}
\includegraphics[scale=0.6]{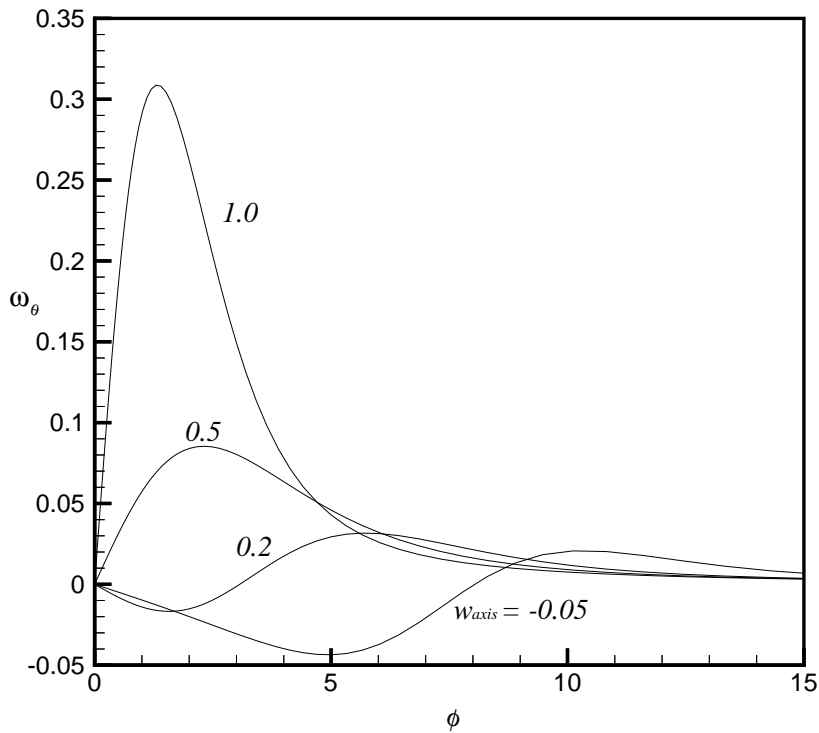}
\vspace{-0.2in}
\caption{The azimuthal vorticity of Long's flow ($n=1$) for $w_{axis}=1.0$. }
\label{n1_ws_wv}
\end{centering}
\end{figure}

In order to obtain at least a preliminary indication as to the hydrodynamic stability of the flow, we begin by observing that the circulation is a monotone increasing function of radius for all of the Long's vortex solutions, so by Rayleigh's criterion (cf.~\cite{Chandrasekhar}, \S66) the flow is stable to inviscid axisymmetric disturbances. The azimuthal vorticity component (figure~\ref{n1_ws_wv}) is everywhere nonnegative for the strictly jet-like flows, but for flows with a wake-like component (i.e.~any region in which $dw/dr=-\omega_\theta>0$) suffers a reversal of sign in the neighborhood of the axis, so the latter solutions satisfy the heuristic criterion of \cite{BrownLopez} for susceptibility to vortex breakdown, at least of the axisymmetric variety.

Even with regard to the jet-like solutions, we know from the scalings in the similarity formulation that conical ($m=-1$) self-similarity implies a severely adverse axial pressure gradient, hence it would be very surprising indeed if such flows were stable (and we know from the extant stability literature that they are not.) More conclusively, we can quickly assure ourselves whether this is so (in the sense of sufficiency, but not necessity) via the stability criterion of \cite{LeibovichStewartson} based on high-azimuthal-wavenumber-mode asymptotics, which applies to generic slender axisymmetric vortex flow at high Reynolds number (the modes in question being subject to rapid viscous damping as Reynolds number decreases) and is defined in terms of our radial similarity coordinate and flow variables as being a sufficient condition for flow instability if the function
\begin{equation}
G_{LS}(\phi)=v\Omega^\p[\Omega^\p g^\p+(w^\p)^2],\qquad{\rm where}\;\Omega=v/r.
\label{LScrit}
\end{equation}
is negative anywhere in the flow field. Plots of the function $G_{LS}(\phi)$  for the the $n=1$ solutions (not shown, owing to the extremely different ranges of variation of this function for the different solutions) show it to always be negative in at least part of the solution domain, which guarantees that the flow is unstable to nonaxisymmetric disturbances as long as the Reynolds number is not small (the latter is required both to assure that the inviscid helical modes involved in the $LS$ criterion are not too heavily damped, and for the assumption of slenderness of the base flow in question to be justified).

Thus, we expect the conically self-similar flows to be unstable, although they still appear to be reasonable from the standpoint of physical realizability (in an idealized disturbance-free world). That is, until we take a careful look at the pressure profiles. The radial distributions of static and total pressure coefficients $C_p$ and $C_{p0}$, defined by~(\ref{pcoeff}), are shown in figures~\ref{n1-Cp-wvar} and~\ref{n1-Cp0-wvar}.

\begin{samepage}

%%% Thesis Fig 5.9, p80
\begin{figure}
\begin{centering}
\includegraphics[scale=0.6]{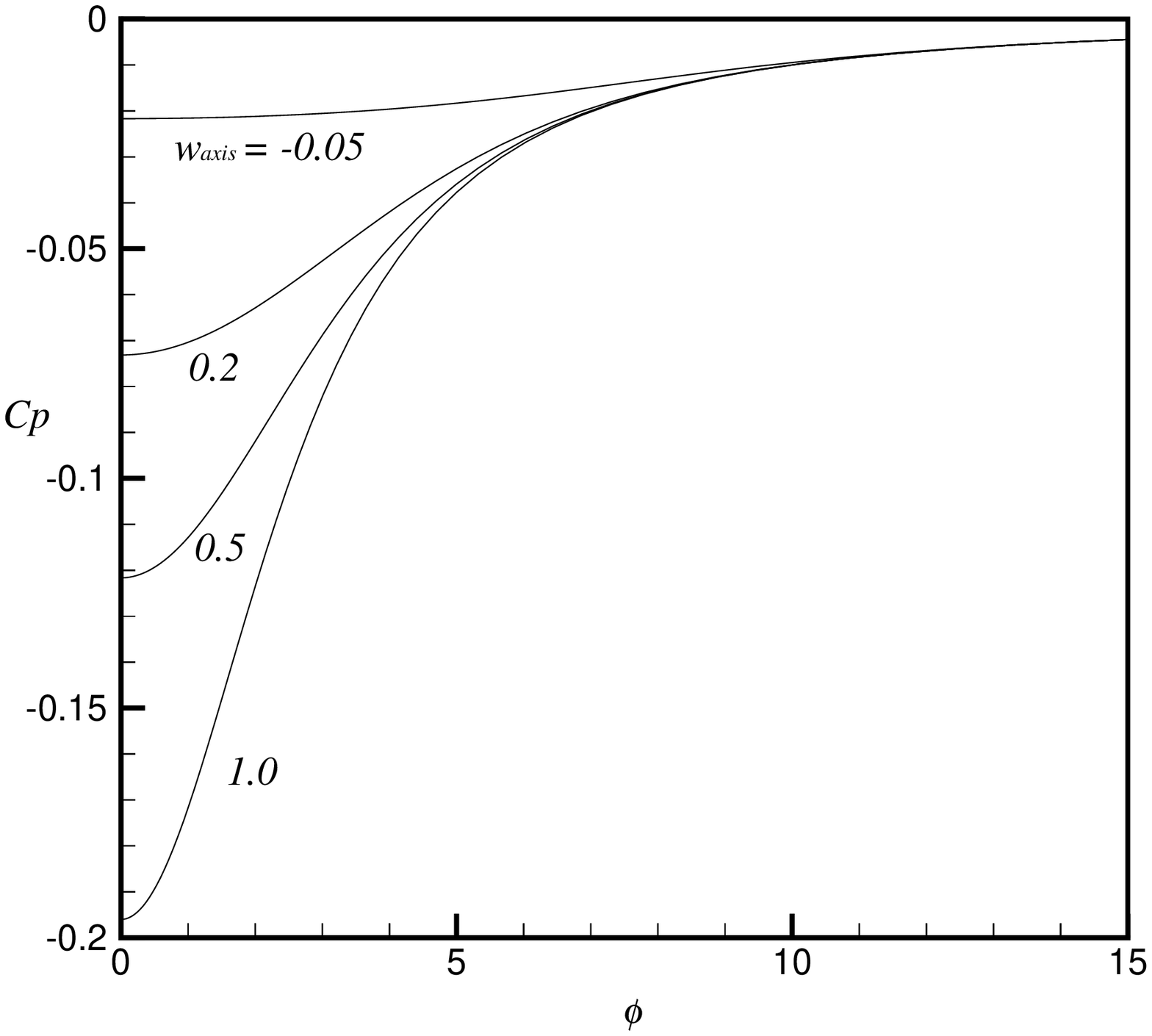}
\vspace{-0.2in}
\caption{The static pressure coefficient of Long's flow ($n=1$) for various $w_{axis}$.}
\label{n1-Cp-wvar}
%\end{centering}
%\end{figure}

%%% Thesis Fig 5.10, p80
%\begin{figure}
%\begin{centering}
\includegraphics[scale=0.6]{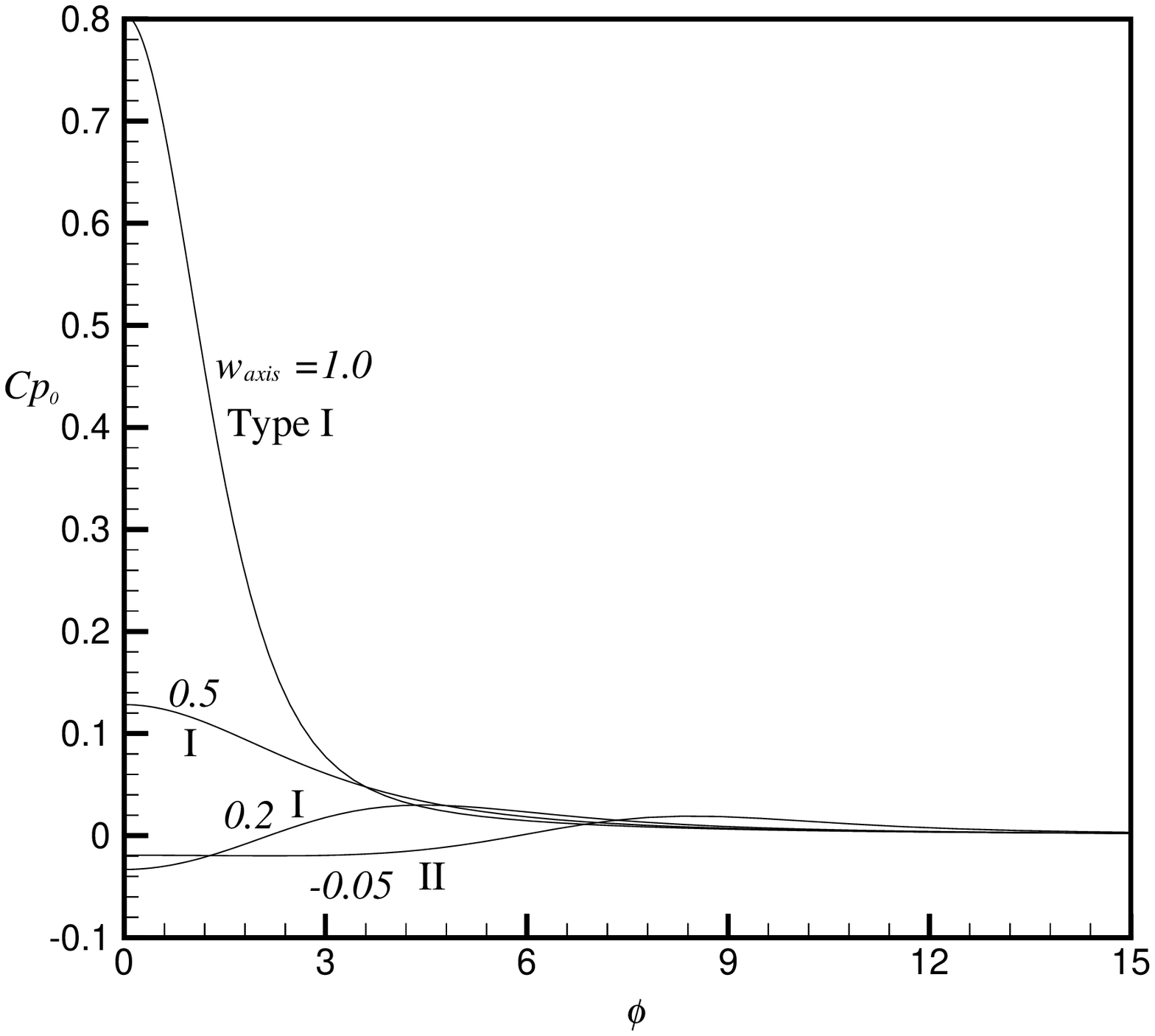}
\vspace{-0.2in}
\caption{The total pressure coefficient of Long's flow ($n=1$) for various $w_{axis}$.}
\label{n1-Cp0-wvar}
\end{centering}
\end{figure}

\end{samepage}

While the static pressure curves all show the expected minimum on the axis, the total  pressure profiles (especially for the jet-like flows) are startling, in that they look like an inverted form of what one expects~-- instead of decreasing in monotone fashion from its far-field value and reaching a minimum on the axis (corresponding to the location where the flow has suffered the most viscous dissipation), $C_{p_0}$ instead reaches a maximum there. We have examined a variety of other well-known viscous vortex flows (viscous time-decaying potential, Burgers, Hall-Stewartson, Mayer-Powell) and observed that Long's is the only one of the lot which shows this type of behavior. Moreover, it is not just a particular subset of the flows which suffers from a nonphysical total-pressure maximum in the viscous region; all of the other conical-flow solutions we have studied, of either Type~I or~II, have clearly nonphysical total-pressure profiles: for the Type~I flows the total-pressure excess tends to occur near the vortex axis, whereas for the Type~II flows it occurs away from the axis. For the large-flow-force asymptotic Type~II flows, the overshoot is confined to the off-axis interior layer, but this different spatial localization gives no reason to believe these flows are any more physical than the others. While this will not be welcome news to those who have invested much time and effort studying the structure and stability of these flows, it seems difficult to believe that such a basic property of the solutions has gone overlooked for over forty years. But we have scoured the extant published literature on the subject and, amazingly, in not a single paper, from the original one by \cite{Long}, to the later work of \cite{BurggrafFoster} and the numerous studies of stability~(cf. \cite{FosterDuck,FosterSmith,FosterJacqmin,KhorramiTrivedi,ArdalanDraperFoster}), have we found a plot of total pressure or mention of the adverse-pressure-gradient nature of these flows.

As it turns out, the  behavior of $C_{p_0}$ can be better understood in the context of the earlier related work of Mayer and Powell, whose flows satisfy the same general similarity laws and equations as those considered here (except for the $n=\frac{1}{2}$ case where there is no axial pressure gradient), albeit with different boundary conditions. They found both favorable and adverse-pressure-gradient flows of a leading-edge-vortex-type character, but observed that solutions satisfying a monotonicity condition on the total pressure could only be found for favorable, neutral, and modestly adverse $\partial p/\partial z$, down to roughly $m=-0.25$
\footnote{
The solutions for $m=-0.4$ plotted in \cite{MayerPowell1} should not have been included, since they actually have a total-pressure overshoot somewhat away from the vortex axis, but the plot which would have shown this was inadvertently excluded from the relevant figure.
}.
For $m$ more negative than this, solutions of  the equations could still be found, but all had near-axis maxima of $C_{p_0}$, the location of the maximum moving radially inward as the pressure gradient became more adverse, until eventually $C_{p_0}$ was strictly monotone {\em decreasing} from the axis outward, much as is the case with the Long's vortex solutions described above. This raises the possibility that there are Long-type solutions of the generalized equations, representing vortical flows in less-severe pressure gradients (and thus having core growth rates less than conical) which perhaps do have physically realistic total-pressure distributions. This indeed proves to be the case, and is the subject of the next section.

%%%%%%%%%%%%%%%%%%%%%%%%%%%%%%
\subsection{Solutions of the General Similarity Equations}
\label{sect:gensolutions}
One of the key numerical quantities associated with the generalized Long-type flows is the leading-order coefficient~$a_{-1}$ of the asymptotic series for the stream function $f$. For a given set of flow parameters $n$ and $w_{axis}$ the value of this term~-- assuming a solution can be found~-- is determined iteratively via the numerical solution procedure described in section~\S\ref{sect:numerics}, and curves of $a_{-1}$ vs.~$n$ for various values of $w_{axis}$ were previously shown in figure~\ref{am1_vs_n}. The most striking feature of the plot is the simultaneous crossing point of all the curves at $n=1$, where $a_{-1}=1/\sqrt{2}$. We are able to obtain well-converged solutions for values of similarity growth-rate parameter ranging from $0.5$ (even though the velocities do not decay at infinity for this value of $n$, the modified functions do and the value of $a_{-1}$ is finite, so we have no trouble finding a numerical solution) to well over unity, that is, for axial pressure gradients ranging from neutral to  even more adverse than those in Long's solutions.

The thesis of Lee has a full variety of plots of the various velocity and vorticity components for the generalized flows with $n$ ranging from just over 0.5 to greater than 1; for the sake of brevity we summarize only the key features here. The axial velocity plots (fig.~\ref{w_w1}) show a striking change as $n$ is varied: for $n$ near one-half, the axial velocity shows a pronounced retardation in the vortex core (which is what one expects physically for an adverse-pressure-gradient flow), but the retardation becomes less pronounced as $n$ increases, and beyond an upper limiting value (which depends on $w_{axis}$, but commonly lies between 0.6 and 0.7), the axial velocity curves revert to the physical implausible jet-like form seen for the Type~I Long's flows. (We illustrate using a slightly wake-like flow with $w_{axis} = -0.07$; the qualitative trend is similar for $w_{axis}=1$ except that the on-axis axial velocity transitions from a local minimum to an outright global maximum as $n$ increases in that case).

\begin{samepage}

%%% Thesis Fig 5.19, p93
\begin{figure}
\begin{centering}
\includegraphics[scale=0.6]{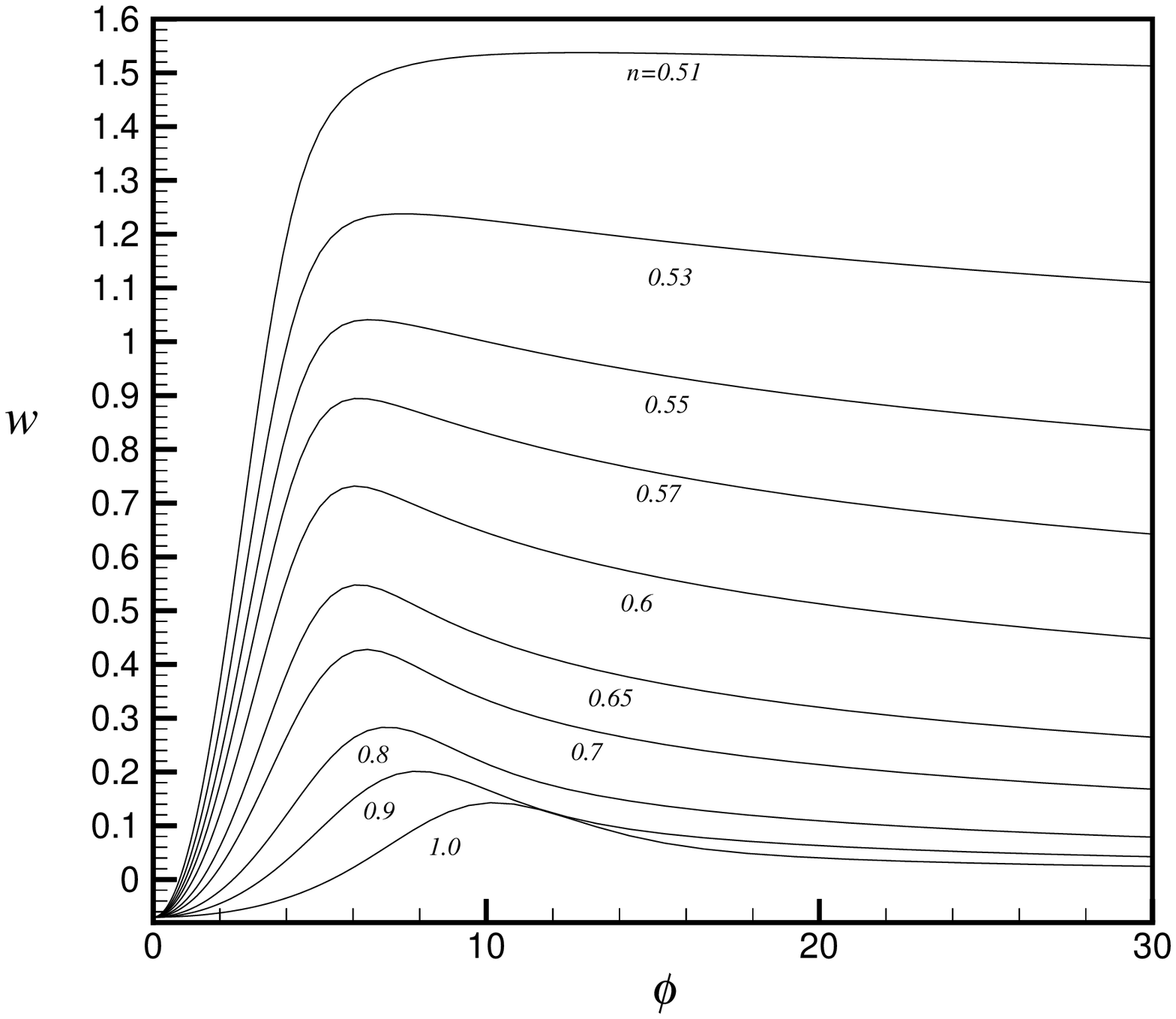}
\vspace{-0.2in}
\caption{Axial velocity distribution of generalized Long's type flows with $w_{axis}=-0.07$ for various values of $n$}
\label{w_w1}
%\end{centering}
%\end{figure}

%%% Thesis Fig 5.20, p93
%\begin{figure}
%\begin{centering}
\includegraphics[scale=0.6]{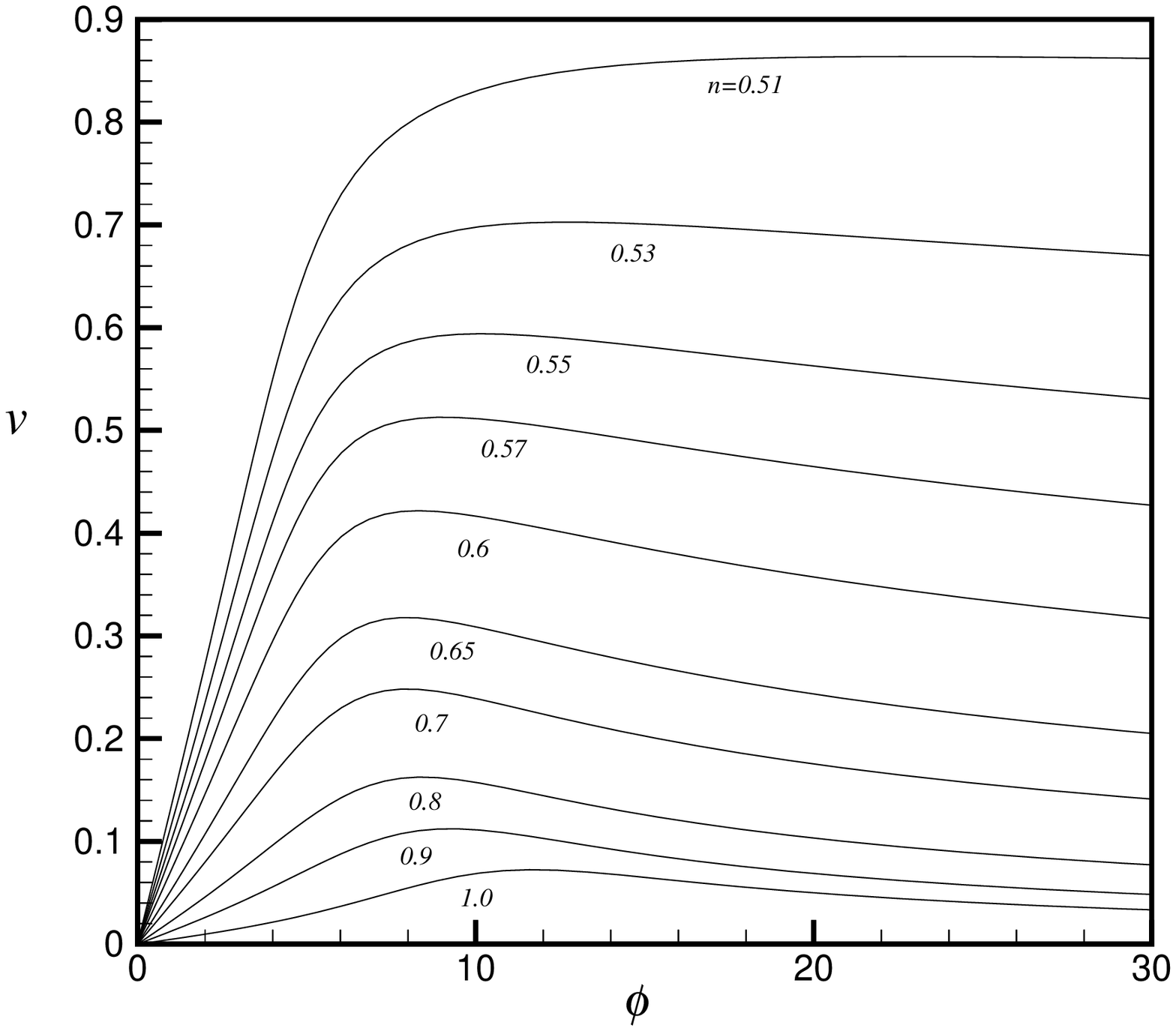}
\vspace{-0.2in}
\caption{Azimuthal velocity distribution of generalized Long's type flows with $w_{axis}=-0.07$ for various values of $n$}
\label{v_w1}
\end{centering}
\end{figure}

\end{samepage}

The radial velocity profiles are broadly similar, but note that the off-axis region where $u$ is negative seen in Long`s case is lacking for modestly adverse pressure gradients. Note that this does not imply a lack of physical fluid entrainment into the vortex core, which would be nonphysical as it would violate conservation of mass; due to the radial growth of the core with downstream distance (and the fact that $u$ is velocity relative to the radial coordinate of a cylindrical polar coordinate system), the ``entrainment velocity'' is not $u$, but rather $u-n\phi w$, and this quantity is indeed everywhere negative except at the axis and at infinity, where it is zero.

The swirl velocity profiles (fig.~\ref{v_w1}) are also qualitatively similar as $n$ is varied, but show two major trends: the maximum value of $v$ increases as $n$ tends to one-half from above (i.e.~as the pressure gradient weakens), and the rate of radial decay decreases. The first property means that swirl-velocity-based measures of vortex strength (which we do not control via boundary conditions, as we do the axial velocity) should take account of this $n$-dependent nature of maximum swirl velocity if one wishes to make quantitative comparisons of the flows at different $n$ based on properties which depend on rotational intensity.

The $u$, $v$  and $w$ velocities all reach their respective extrema at order-unity values of $\phi$ before decaying further out. As expected from the asymptotic analysis, the ${\cal O}(\phi^{-1})$ decay rate of $u$ is faster than the ${\cal O}(\phi^{\frac{1}{n}-2})$ decay of $v$ and $w$ for $n<1$.

%The axial vorticity, $\omega_z=v^\prime-v/\phi$ (fig.~\ref{omega_z_w1}) mirrors this trend: as $n\rightarrow0.5^{+}$, the peak value of $\omega_z$ increases, and the vorticity decays more slowly in the far field. Profiles of the azimuthal vorticity $\omega_{\theta}$ are shown in figures~\ref{omega_theta_w1}. Azimuthal vorticity becomes negative near the rotation axis as $n$ decreases from 1 and approaches the ``physical region''; these negative $\omega_{\theta}$ values correlate with the retardation of axial velocity near the vortex axis. Smaller values of $n$ correspond to more negative values of $\omega_{\theta}$ near the axis, which seems counterintuitive at first glance, but owing to the different underlying swirling velocities (and thus vortex strengths, roughly speaking), it is not the magnitude of $\omega_{\theta}$, but the sign that is meaningful when comparing solutions with fixed $w_{axis}$ and different $n$.

\begin{figure}
\begin{centering}
\includegraphics[scale=0.7]{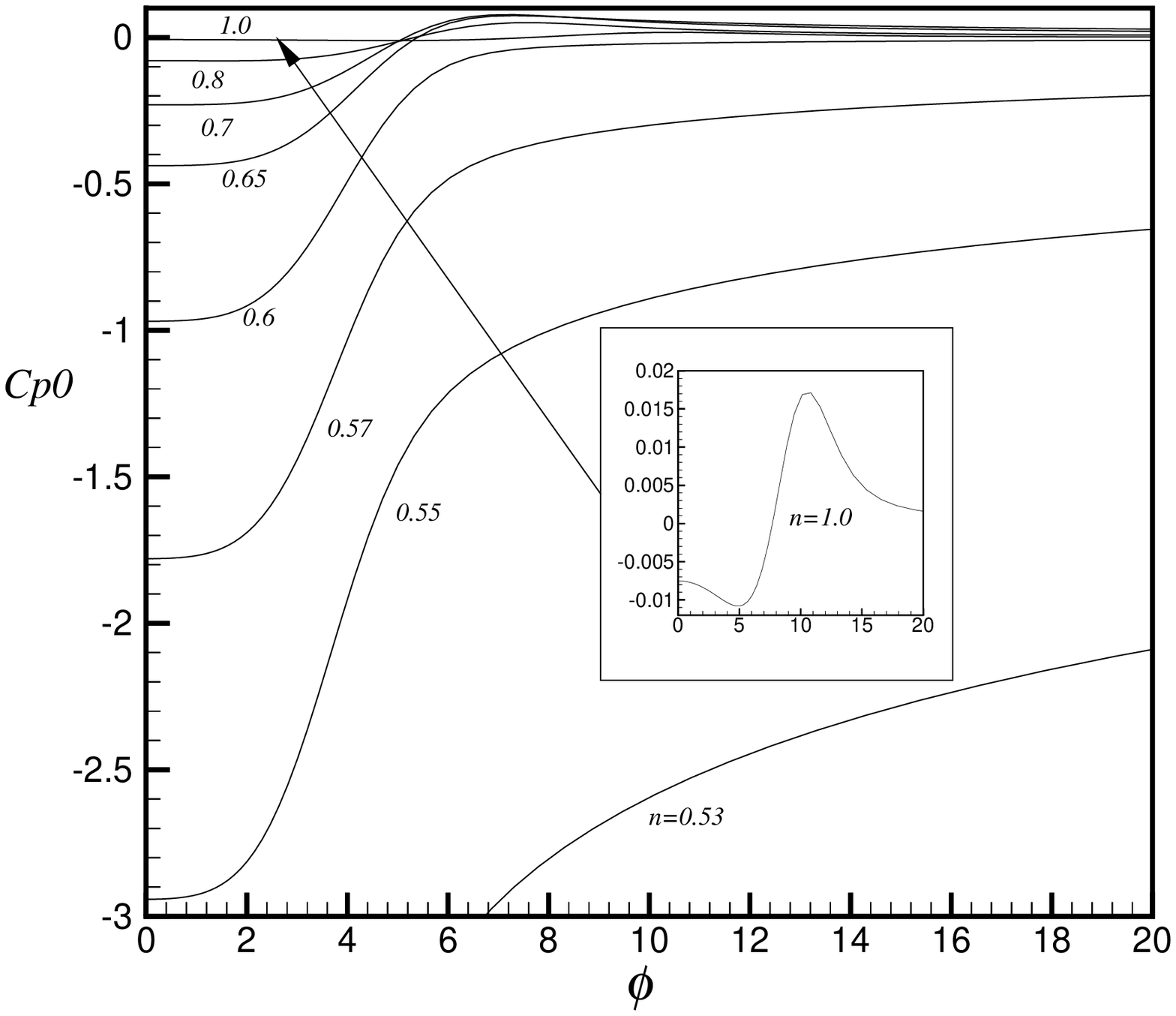}
\vspace{-0.2in}
\caption{Distribution of $C_{p_0}$ of generalized Long's type flows with $w_{axis}=-0.07$ for various values of $n$}
\label{cp0_w1}
\end{centering}
\end{figure}

As to the physicality of the generalized flows, total pressure coefficient distributions, again all for $w_{axis}=-0.07$, appear in figure~\ref{cp0_w1}. (The case $n=1$ and  $w_{axis}=1$ was already shown in the discussion on physicality of Long's flows). For $n$ near one-half, the profiles look quite reasonable: a pronounced deficit in the vortex core, and tending to a constant maximum in the far field. As $n$ increases a small off-axis overshoot appears, and for larger $n$ or larger values of $w_{axis}$ the overshoots become worse, eventually turning into the completely inverted-looking profiles already seen for Long's case. Note that owing to the different underlying swirling velocities (and thus vortex strengths, roughly speaking), it is not the absolute magnitude of total-pressure coefficient but the sign that is meaningful when comparing solutions with fixed $w_{axis}$ and different $n$.

%%% Thesis Fig 5.23, p95

From the cases we calculated, we list in Table~\ref{physicality} the range of $n$ in which the total pressure coefficient satisfies the physicality criterion for various central axial velocities. We find that in general, higher values of $w_{axis}$ yield narrower ranges of physicality. Since the values of this parameter is specified as a boundary condition, such a trend implies that the jet velocity a physically reasonable flow can attain under a more adverse axial pressure gradient is smaller than for flows subject to less adverse pressure gradients, a physically intuitive result. The lower limit of the range of $w_{axis}$, namely $-0.07$, is about the lowest $w_{axis}$ for which we can find converged solutions using the previously described numerical scheme, at least without making any special allowance in the computational mesh for the increasingly ring-like nature of the flow. On the other hand, one can find converged solutions for much higher $w_{axis}$ than $2.0$ easily. As the character of these high $w_{axis}$ flows is qualitatively similar to $w_{axis}=1.0$ ones, we do not show them here.

\begin{table}
\vspace{0.5cm}
\begin{center}
\begin{tabular}{c|c|c|c|c|c|c|c|c}
$w_{axis}$&2.00&1.50&1.0&0.75&0.50&0.20&$-$0.01&$-$0.07\\ \hline
$n_{max}$ of Physicality&0.5425&0.558&0.58&0.59&0.60&0.61&0.60&0.60
\end{tabular}
\end{center}
\vspace{-0.2in}
\caption{The range of physicality for each case is $0.5<n\le n_{max}$}
\label{physicality}
\end{table}

As far as the stability of the generalized Long-type flows is concerned, the preliminary (in the sense of a strictly temporal linear normal-mode analysis) results presented in \cite{LeeThesis} indicate a broadly similar spectrum of unstable bending-wave and nonaxisymmetric invisicid modes as has been well-documented for the conical case by \cite{FosterDuck}, \cite{FosterSmith}, and \cite{ArdalanDraperFoster}. Unlike the conical case studies by \cite{KhorramiTrivedi}, we do find some weakly unstable viscous modes for the generalized flow solutions.

%%%%%%%%%%%%%%%%%%%%%%%%%%%%%%
\subsection{Relationship Between Long and Hall-type Solutions}
\label{sect:long2hall}

By way of conclusion, we shall explicitly demonstrate the relation between the generalized Long-type flows, the generalized Hall-type flows studied by Mayer~\&~Powell, and the edge pressure of the flow considered in a finite domain $[0,\phi_e]$. As mentioned in section~\ref{sect:intro}, for flows with nonzero axial pressure gradient ($m\ne 0$) the edge pressure at a finite value of $\phi_{e}$ has a deterministic effect on the flow solution. In fact, one can make a transition from a generalized Long type flow to a Hall type like flow by adjusting the edge pressure $p_e$. To demonstrate this, we first calculate the solution for a case of generalized Long-type flow in an infinite domain, then pick a finite value of $\phi_{e}$ and cut off solutions beyond it. We thus have the velocity and pressure fields within $[0,\phi_e]$ of a Long-type flow, but finite-radius outer boundary conditions of the kind used to generate Hall-type solutions. Now by keeping $v_e$ and $w_e$ fixed at $\phi_e$ and adjusting $p_e$, we can solve for the flow field within $[0,\phi_e]$ in the same way \cite{MayerPowell1} solved for the generalized Hall-type flows, and for the proper value of $p_e$ obtain the same solution as the latter authors. This transition behavior is most pronounced for the axial component of velocity and for the total pressure, which are shown in figures~\ref{long2hall_w} and~\ref{long2hall_cp0}, for a modestly-adverse-pressure-gradient scenario (i.e.~one for which both a Long-type and Hall-type solution exists) having $n=0.6$. The uppermost curve of each set shows the distribution of the flow quantity in question for the Long-type solution at the chosen arameter values. The successively-lower curves represent the effects of gradually lowering the finite-outer-boundary pressure value toward the value of $p_e=\frac{1}{2}(v_e^2+w_e^2)$ of the Hall-type flows. As the plots show, as $p_e$ is lowered the axial velocity distribution transitions from the nearly pure jet-like profile of the Long-type solution to ones exhibiting the pronounced retardation at the axis already familiar from the generalized Hall-type solutions. The large set of curves in the figure shows the precise $p_e$ value associated with each curve; the smaller inset shows the same curves plotted all the way out to the chosen outer-boundary location $\phi_e=40$.

\begin{samepage}

%%% Thesis Fig x.y:
\begin{figure}
\begin{centering}
\includegraphics[scale=0.6]{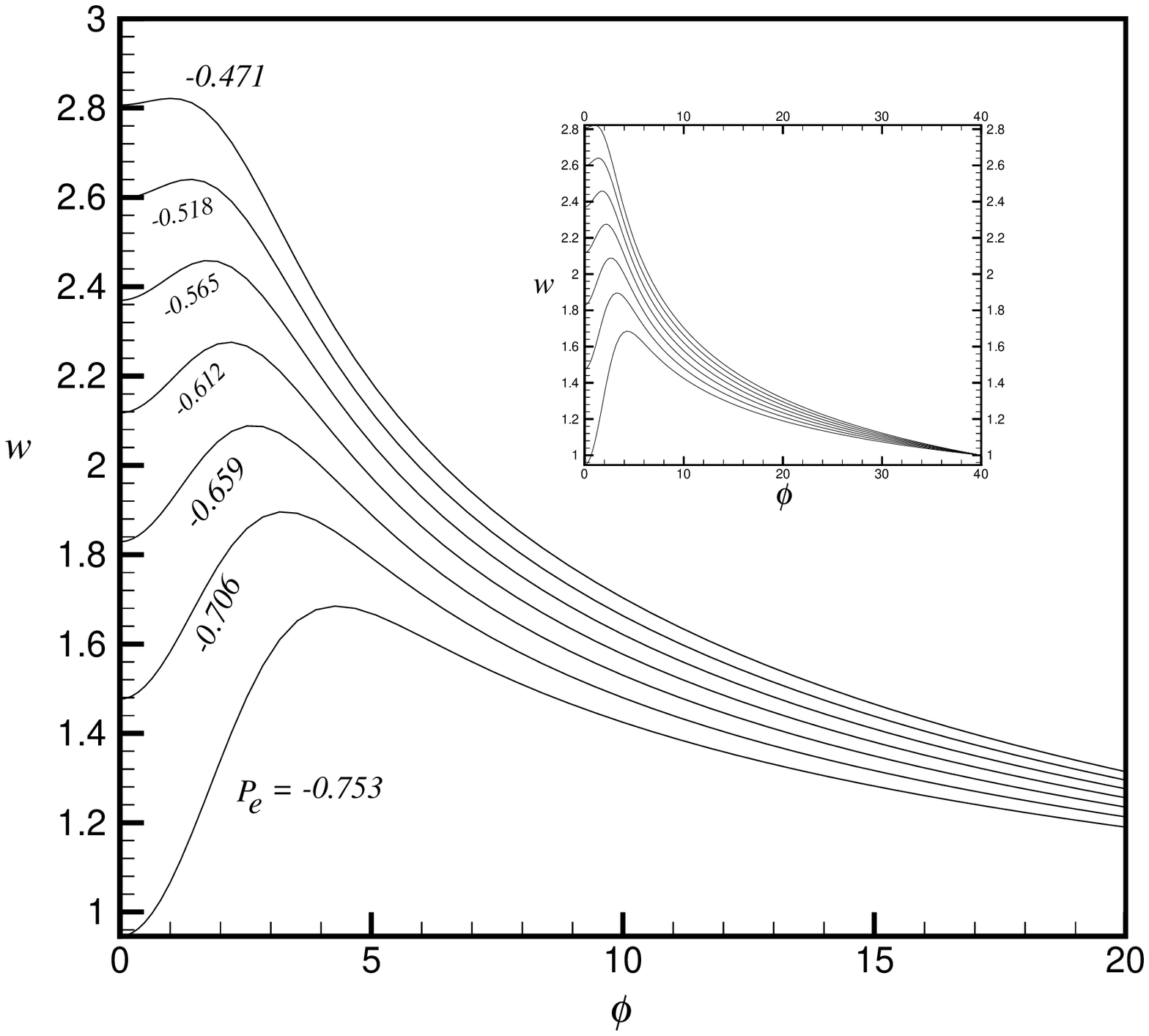}
\vspace{-0.2in}
\caption{The transition of velocity $w$ from a generalized Long-type flow to a Hall-type-like flow  for $n=0.6$.}
\label{long2hall_w}
%\end{centering}
%\end{figure}

%%% Thesis Fig x.y:
%\begin{figure}
%\begin{centering}
\includegraphics[scale=0.6]{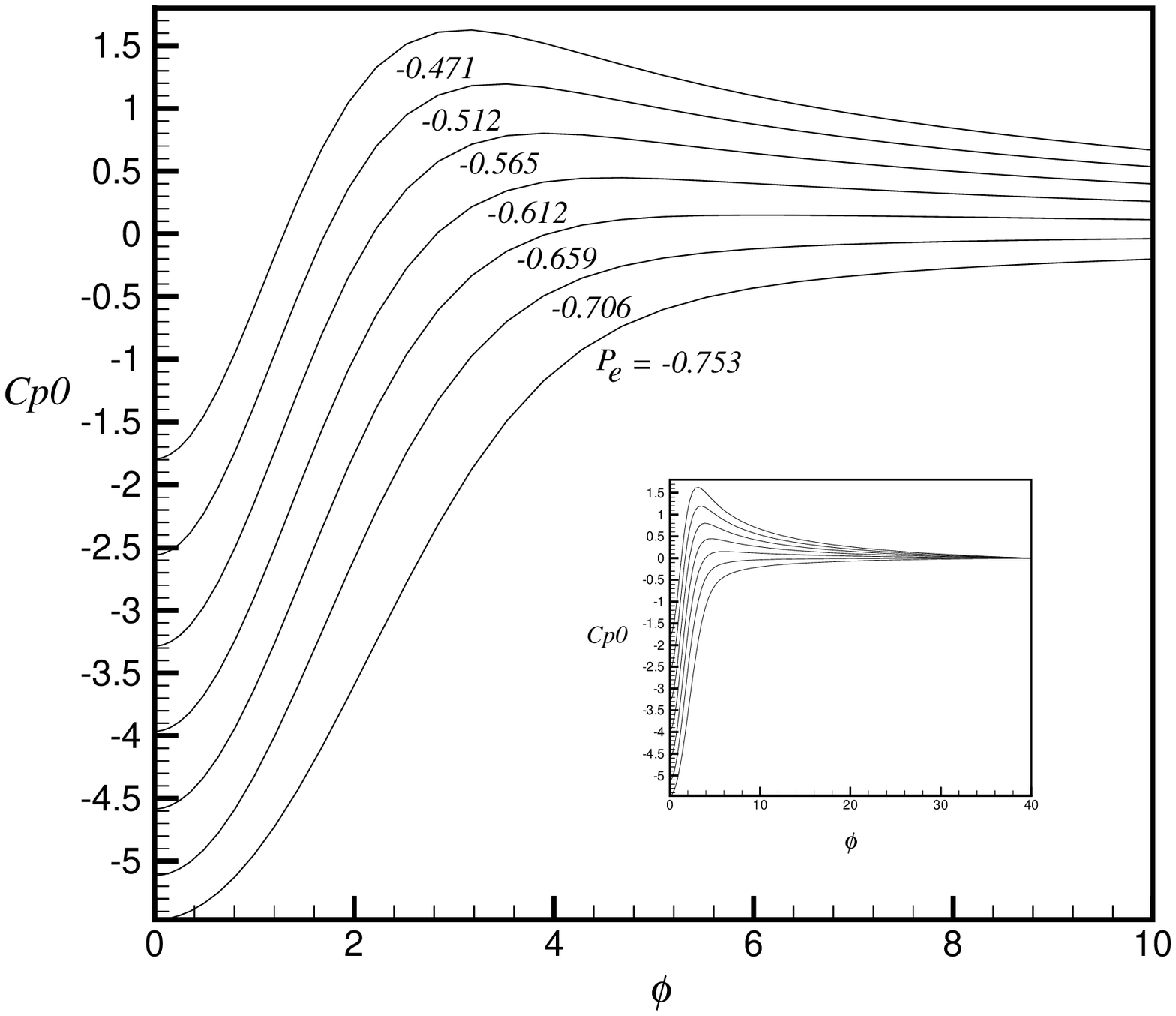}
\vspace{-0.2in}
\caption{The transition of total pressure coefficient $C_{p_0}$ from a generalized Long-type flow (nonphysical) to a Hall-type-like flow (physical) for $n=0.6$.}
\label{long2hall_cp0}
\end{centering}
\end{figure}

\end{samepage}

The total-pressure-coefficient plots show the transition from the Long-type ones having a pronounced overshoot near the axis~-- just how pronounced is clearer in the inset~-- to the monotone increasing profile of the Hall-type flow. As was summarized in Table~\ref{physicality}, there are in fact (only just) some Long-type flows at $n=0.6$ which possess monotone total-pressure profiles without requring such surgical intervention, but they exist at much smaller values of axial velocity than the cases shown here.

%%%%%%%%%%%%%%%%%%%%%%%%%%%%%%%
\section{Acknowledgements}
\label{sect:acks}

The authors would like to thank Professor S.~N.~Brown (University College London) for helpful discussions during the early stages of this work, and Professor K.~G.~Powell (University of Michigan, Ann Arbor) for providing us with a copy of the paper of Morton.

\clearpage

%%%%%%%%%%%%%%%%%%%%%%%%%%%%%%%

% Bibliography
\bibliographystyle{acmsmall}
\bibliography{LongsVortexRevisited}

\end{document}